\documentclass[11pt]{article}
\usepackage[dvips]{graphicx,psfrag}

\begin{document}
\title{New CP Phase and Exact Oscillation Probabilities \\of Dirac Neutrino 
derived from Relativistic Equation}
\author{
{K. Kimura$^1$}
{and A. Takamura$^{2}$}
\\
\\
\\
{\small \it $^1$Department of Physics, Nagoya University,}
{\small \it Furo-cho, Chikusa-ku, Nagoya, 464-8602, Japan}\\
{\small \it $^2$Department of Mathematics, 
Toyota National Collage of Technology}\\
{\small \it Eisei-cho 2-1, Toyota-shi, 471-8525, Japan}}
\date{}
\maketitle
\vspace{-11cm}
\begin{flushright}
\end{flushright}
\vspace{10.5cm}
\vspace{-2.5cm}
%
%
\vspace{1cm}
\begin{abstract}
We present a new formulation for deriving neutrino oscillation probabilities relativistically, 
based not on the Schrödinger equation but on the Dirac equation. In the context of two generations, we calculate the oscillation probabilities precisely in a scenario where only the Dirac mass term is present. 
Our analysis reveals the emergence of two new features in the oscillation probabilities derived from the Dirac equation.
The first feature is that the oscillation probabilities depend on the absolute value of the neutrino mass.
While it has generally been assumed that oscillation probabilities depend solely on mass squared differences, we show that they also depend on the absolute mass.
The second feature is the emergence of a new CP phase.
If interactions exist that can distinguish the flavors of right-handed neutrinos 
in physics beyond the Standard Model, 
we could potentially observe this new CP phase even in the two-generation model.
We discuss the feasibility of detecting the contributions of these features through neutrino oscillations at atomic scales. 
In contrast, these effects are negligible in conventional short- and long-baseline experiments, 
and no contradictions arise with previous findings.
\end{abstract}



\section{Introduction}
\label{sec:intro}

In 1957, Pontecorvo proposed the concept of neutrino-antineutrino oscillations, drawing inspiration from K$^0$-$\bar{\rm K}^0$ oscillations in the presence of lepton number violation \cite{Pontecorvo}. Subsequently, in 1962, following the discovery of muon neutrinos, Maki, Nakagawa, and Sakata introduced the idea of oscillations among neutrinos of different flavors \cite{MNS}. Since then, various theories of neutrino oscillations have been developed, including those involving three generations \cite{3-gene}, matter effects \cite{MSW1, MSW2}, magnetic fields \cite{mag}, and non-standard interactions \cite{grossman, NSI1, NSI2, NSI3, NSI4}.

The phenomenon of neutrino oscillation was experimentally confirmed in 1998 by the atmospheric neutrino experiment conducted at Super-Kamiokande \cite{SKatmos}. Following this discovery, further evidence for neutrino oscillations has been accumulated from solar neutrino experiments \cite{SK, SNO, SK2}, long-baseline experiments \cite{T2K, MINOS, NOvA}, and reactor experiments \cite{KamLAND, DayaBay, RENO, DoubleChooz}. Analyses of these experiments have clarified the values of two mass-squared differences and three mixing angles. Notably, recent results from the T2K experiment suggest that the value of the Dirac CP phase is close to that of maximal CP violation \cite{Dirac CP}. Next-generation experiments \cite{HK, DUNE} are planned to investigate the CP phase and neutrino mass ordering. For these experiments, it is crucial to estimate matter effects as accurately as possible for precise measurements of neutrino parameters. To this end, the theory of neutrino oscillation incorporating matter effects has been rigorously formulated using the Schrödinger equation \cite{Zaglauer, Ohlsson, KTY1, KTY2, Yokomakura0207, Yasuda}.

The oscillations confirmed by prior experiments all involve neutrinos without chirality-flip. However, oscillations between neutrinos of different chiralities have also been proposed. The Dirac equation allows for the possibility of chirality change, as both $\nu_L$ and $\nu_R$ are included in the same multiplet. In the context of one-generation Dirac neutrinos, Fukugita and Yanagida calculated the probability of $\nu_L \to \nu_R$ oscillation \cite{Fukugita-note}. Similarly, in the case of Majorana neutrinos \cite{Majorana}, oscillation probabilities between neutrinos and antineutrinos have been derived in both two- and three-generation frameworks \cite{Bahcall1978, Valle1981, Li1982, Bernabeu1983, Gouvea2003, Xing2013}. These also represent oscillations between neutrinos with different chiralities.

For oscillations involving chirality-flip, the probabilities are proportional to $m^2/E^2$, where $E$ is the neutrino energy and 
$m$ is its mass. As a result, measuring this small effect is considered challenging. Nonetheless, if oscillations with chirality-flip coexist alongside established oscillations among neutrinos of the same chirality, the sum of the probabilities for 
$\nu_{eL}\to\nu_{eL}$, $\nu_{eL} \to \nu_{\mu L}$ and $\nu_{eL} \to \nu_{\tau L}$ would not equal one, implying a violation of unitarity. This stems from the consideration of only left-handed neutrinos within the framework of the Schrödinger equation.

In this paper, we formulate the theory of neutrino oscillations using the relativistic Dirac equation, aiming to unify cases with and without chirality-flip. We focus on the scenario of two-generation neutrinos with a Dirac mass term in vacuum. The results obtained can be extended to include Majorana neutrinos, $n$ generations, and cases involving matter effects or magnetic fields. We demonstrate that unitarity holds when considering both types of oscillations.

Our analysis reveals the emergence of two new features in the derived probabilities: one is dependent on the absolute mass of neutrinos, and the other includes a new CP phase. This new CP phase manifests in oscillations with chirality-flip, and the Majorana CP phase can be interpreted as one example of this phase.

Consequently, we show that it is possible, in principle, to measure not only mass-squared differences but also the absolute mass of neutrinos through oscillation experiments. Furthermore, the new CP phase could be detectable if interactions exist that can distinguish the flavors of right-handed neutrinos, even within the context of two generations. The contributions of these new fearures may yield non-negligible effects in atomic-scale experiments. Conversely, in conventional short, medium, and long-baseline experiments, their contributions are negligible, thus aligning with previous results.

The structure of the paper is organized as follows: In Section II, we define the notation used throughout this work. Section III reviews the derivation of neutrino oscillation probabilities as developed in previous papers using the Schrödinger equation. In Section IV, we present the relativistic derivation of neutrino oscillation probabilities from the Dirac equation and verify unitarity. Finally, Section V summarizes the results obtained in this paper.

\section{Notation}

In this section, we write down the notation used in this paper. 
We mainly use the chiral representation 
because neutrinos are measured through weak interactions.
In chiral representation, the gamma matrices with $4\times 4$ form are given by 
\begin{eqnarray}
\gamma^0=\left(\begin{array}{cc}0 & 1 \\ 1 & 0\end{array}\right), \qquad 
\gamma^i=\left(\begin{array}{cc}0 & -\sigma_i \\ \sigma_i & 0\end{array}\right), \qquad 
\gamma_5=\left(\begin{array}{cc}1 & 0 \\ 0 & -1\end{array}\right),  \label{gamma-mat}
\end{eqnarray}
where $2\times 2$ $\sigma$ matrices are defined by 
\begin{eqnarray}
\sigma_1=\left(\begin{array}{cc}0 & 1 \\ 1 & 0\end{array}\right), \qquad 
\sigma_2=\left(\begin{array}{cc}0 & -i \\ i & 0\end{array}\right), \qquad 
\sigma_3=\left(\begin{array}{cc}1 & 0 \\ 0 & -1\end{array}\right). 
\end{eqnarray}
We also define 4-component spinors $\psi$, $\psi_L$ and $\psi_R$ as 
\begin{eqnarray}
\psi=\left(\begin{array}{c}\xi \\ \eta \end{array}\right), \qquad 
\psi_L=\frac{1-\gamma_5}{2}\psi=\left(\begin{array}{c}0 \\ \eta\end{array}\right), \qquad 
\psi_R=\frac{1+\gamma_5}{2}\psi=\left(\begin{array}{c}\xi \\ 0 \end{array}\right),   \label{psi-def}
\end{eqnarray}
and 2-component spinors $\xi$ and $\eta$ as 
\begin{eqnarray}
\xi=\left(\begin{array}{c}\nu_R^{\prime} \\ \nu_R \end{array}\right), \qquad 
\eta=\left(\begin{array}{c}\nu_L^{\prime} \\ \nu_L \end{array}\right).  \label{xi-eta}
\end{eqnarray}

We consider the quantum numbers of each component that appears in (\ref{xi-eta}).
We use three operators to distinguish the states of neutrinos: chirality, helicity, and energy.
We calculate the commutation relation between the chirality operator $\gamma_5$, the helicity operator $\Sigma\cdot \hat{p}$, and the Hamiltonian $H$. Here, $\Sigma$ is a matrix arranged in diagonal components with two Pauli matrices, $\hat{p}$ is a unit vector with momentum direction, and
we take the momentum in the $z$ direction.
First, the commutation relation between $\gamma_5$ and $H$ is
\begin{eqnarray}
[\gamma_5, H]&=&[\gamma_5, {\bf \alpha}\cdot p+\beta m]=[\gamma_5, \gamma^0\gamma^i p_i+\gamma^0 m]. 
\nonumber \\\
&=&\gamma_5(\gamma^0\gamma^i p_i+\gamma^0 m)-(\gamma^0\gamma^i p_i+\gamma^0 m)\gamma_5
=2\gamma_5\gamma^0 m\neq 0, 
\end{eqnarray} 
and does not commute. 
This means that the left-handed neutrino produced as an eigenstate of chirality can change over time and become a right-handed neutrino.

Second, the commutation relation between $\gamma_5$ and the helicity operator $\Sigma\cdot \hat{p}$ is
\begin{eqnarray}
[\gamma_5, \Sigma\cdot \hat{p}]&=&\left(\begin{array}{cc}
1 & 0 \\
0 & -1 
\end{array}\right)
\left(\begin{array}{cc}
\sigma\cdot \hat{p} & 0 \\ 
0 & \sigma\cdot \hat{p} 
\end{array}\right)-\left(\begin{array}{cc}
\sigma\cdot \hat{p} & 0 \\ 
0 & \sigma\cdot \hat{p} 
\end{array}\right)\left(\begin{array}{cc}
1 & 0 \\
0 & -1 
\end{array}\right)
=0
\end{eqnarray} 
This means that chirality and helicity can be diagonalized simultaneously. 

Finally, the commutation relation between helicity and Hamiltonian is well known as
\begin{eqnarray}
[H, \Sigma\cdot \hat{p}]&=&
\left[
\left(\begin{array}{cc}
\sigma\cdot p & 0 \\ 
0 & -\sigma\cdot p 
\end{array}\right)+\left(\begin{array}{cc}
0 & m \\
m & 0 
\end{array}\right),
\left(\begin{array}{cc}
\sigma\cdot \hat{p} & 0 \\ 
0 & \sigma\cdot \hat{p} 
\end{array}\right)\right]
=0
\end{eqnarray}
This means that the helicity is conserved and can be diagonalized simultaneously with the Hamiltonian. 
This is also consistent with the conservation law of angular momentum.

Taking into account that neutrinos are produced through weak interactions, we take the initial state of the neutrino to be an eigenstate of chirality and helicity.
Since neutrinos have a very small but finite mass, there are states with both positive and negative helicity 
for each neutrino with left-handed and right-handed chirality.
For neutrinos produced with certain chirality and helicity, the chirality changes with time, but its helicity does not.
Also note that in this case, the initial state of the neutrino is not an energy eigenstate.

As we can see by multiplying the chirality and helicity operators
to the neutrino state, the components, $\nu_L$ and $\nu_R$ in (\ref{xi-eta}), are states with negative helicity, as shown in the Table 1 , and $\nu_L^{\prime}$ and $\nu_R^{\prime}$ have positive helicity.
Also, antineutrinos are defined as charge conjugation of neutrinos, so their momentum is opposite to that of neutrinos.
For this reason, the helicity of antineutrinos is of the opposite sign to that of neutrinos, even though their spin direction is the same.

\begin{table}[tb]
  \begin{center} 
    \caption{chirality and helicity of each component}
\begin{tabular}{|c|c|c|c|c|c|c|c|c|}
\hline
& $\nu_R^{\prime}$ & $\nu_R$ & $\nu_L^{\prime}$ & $\nu_L$ 
& $\nu_R^{c\prime}$ & $\nu_R^c$ & $\nu_L^{c\prime}$ & $\nu_L^c$\\\hline
chirality & Right & Right & Left & Left & Left & Left & Right & Right \\\hline
helicity & $+$ & $-$ & $+$ & $-$ & $-$ & $+$ & $-$ & $+$
\\\hline
\end{tabular}
  \end{center}
\end{table}

Furthermore, 
we use the subscript $\alpha$ and $\beta$ for flavor, $L$ and $R$ for chirality, 
the number $1$ and $2$ for generation and superscript $\pm$ for energy. 
Because of negligible neutrino mass, mass eigenstate has been often identified with energy eigenstate 
in many papers.
But in the future, we should distinguish these two kinds of eigenstates because of the finite neutrino mass. 
More concretely, we use the following eigenstates; 
\begin{eqnarray}
&&{\rm flavor \,\, eigenstates}: \quad \nu_{\alpha L}, \nu_{\alpha R}, \nu_{\beta L}, \nu_{\beta R}, \\
&&{\rm mass \,\, eigenstates}: \quad\,\, \nu_{1L}, \nu_{1R}, \nu_{2L}, \nu_{2R}, \\
&&{\rm energy \,\, eigenstates}: \quad  \,\,\,\nu_1^+, \nu_1^-, \nu_2^+, \nu_2^-.
\end{eqnarray}
It is noted that mass eigenstates are not exactly the eigenstates.
We use the term, eigenstates, in the sense that the mass submatrix in the Hamiltonian is diagonalized.
We also difine the spinors for antineutrino as charge conjugation of neutrino $\psi^c=i\gamma^2 \psi^*$.
The charge conjugations for left-handed and right-handed neutrinos are defined by 
\begin{eqnarray}
\psi_L^c\equiv (\psi_L)^c\!\!\equiv\!\!
\left(\begin{array}{c}\nu_L^c \\ \nu_L^{c\prime} \\ 0 \\ 0 \end{array}\right) 
\!\!\equiv i\gamma^2 \psi_L^*\!=\!i\gamma^2 \frac{1-\gamma_5}{2}\psi^*\! 
=\!\frac{1+\gamma_5}{2}(i\gamma^2 \psi^*)
\!=\!\!(\psi^c)_R\!\!=\!\left(\!\!\begin{array}{c}i\sigma_2 \eta^* \\ 0 \end{array}\!\!\right)
\!=\!\left(\!\!\begin{array}{c}\nu_L^* \\ -\nu_L^{*\prime} \\ 0 \\ 0 \end{array}\!\!\right), \label{nuc} 
\end{eqnarray}
\begin{eqnarray}
\psi_R^c\equiv (\psi_R)^c\!\!\equiv\!\!\left(\begin{array}{c}0 \\ 0 \\ \nu_R^c \\ \nu_R^{c\prime} \end{array}\right)
\!\!\equiv i\gamma^2 \psi_R^*\!=\!i\gamma^2 \frac{1+\gamma_5}{2}\psi^*\! 
=\!\frac{1-\gamma_5}{2}(i\gamma^2 \psi^*)
\!=\!\!(\psi^c)_L\!\!=\!\left(\!\!\begin{array}{c}0 \\ -i\sigma_2 \xi^* \end{array}\!\!\right)
\!=\!\left(\!\!\begin{array}{c}0 \\ 0 \\ -\nu_R^* \\ \nu_R^{*\prime} \end{array}\!\!\right). 
\end{eqnarray}
It is noted that the chirality is flipped by taking the charge conjugation.

\section{Conventional Derivation of Oscillation Probabilities}

In this section, we consider the case of Dirac neutrinos 
and review the derivation of the oscillation probabilities performed in previous papers. \\

\subsection{Conventional Derivation of Chirality non-flipped Probabilities}

At first, we review how the oscillation probabilities without chirality-flip are derived by the Schr$\ddot{\rm o}$dinger equation.
In vacuum, 
time evolution of flavor eigenstates (weak eigenstates), $\nu_{\alpha L}$ and $\nu_{\beta L}$ 
after the time $t$, is given by  
\begin{eqnarray}
\left(\begin{array}{c}
\nu_{\alpha L}(t) \\ \nu_{\beta L}(t) 
\end{array}\right)
=\left(\begin{array}{cc}
U_{\alpha 1} & U_{\alpha 2} \\ 
U_{\beta 1} & U_{\beta 2} 
\end{array}\right)\left(\begin{array}{c}
\nu_{1}^+(t) \\ \nu_{2}^+(t) 
\end{array}\right)
=\left(\begin{array}{cc}
U_{\alpha 1} & U_{\alpha 2} \\ 
U_{\beta 1} & U_{\beta 2} 
\end{array}\right)\left(\begin{array}{cc}
e^{-iE_1 t} & 0 \\ 
0 & e^{-iE_2 t} 
\end{array}\right)\left(\begin{array}{c}
\nu_{1}^+ \\ \nu_{2}^+ 
\end{array}\right),
\end{eqnarray}
where $U_{\alpha i}$ and $U_{\beta i}$ are the elements of unitary matrix, $E_i$ is the energy of $\nu_i^+$ 
and $\nu_i^+$ means $\nu_i^+(0)$. 
It is noted that the energy eigenstates are identified with so-called mass eigenstates in conventional derivation.
Rewriting the relations of the fields to those of the one particle states, we obtain 
\begin{eqnarray}
|\nu_{\alpha L}(t)\rangle &=&U_{\alpha 1}^*e^{-iE_1t}|\nu_1^+\rangle+U_{\alpha 2}^*e^{-iE_2t}|\nu_2^+\rangle, \\
|\nu_{\beta L}(t)\rangle &=&U_{\beta 1}^*e^{-iE_1t}|\nu_1^+\rangle+U_{\beta 2}^*e^{-iE_2t}|\nu_2^+\rangle, 
\end{eqnarray}
and the conjugate states 
\begin{eqnarray}
\langle \nu_{\alpha L}| &=&U_{\alpha 1}\langle \nu_1^+|+U_{\alpha 2}\langle \nu_2^+|, \\
\langle \nu_{\beta L}| &=&U_{\beta 1}\langle \nu_1^+|+U_{\beta 2}\langle \nu_2^+|. 
\end{eqnarray}
Therefore, we have the transition amplitudes of $\nu_{\alpha L}$ to $\nu_{\alpha L}$ and $\nu_{\beta L}$, 
\begin{eqnarray}
A(\nu_{\alpha L}\to\nu_{\alpha L})&=&\langle \nu_{\alpha L}|\nu_{\alpha L}(t)\rangle =
|U_{\alpha 1}|^2e^{-iE_1t}+|U_{\alpha 2}|^2e^{-iE_2t}, \label{con-am1}\\
A(\nu_{\alpha L}\to\nu_{\beta L})&=&\langle \nu_{\beta L}|\nu_{\alpha L}(t)\rangle 
=U_{\alpha 1}^*U_{\beta 1}e^{-iE_1t}+U_{\alpha 2}^*U_{\beta 2}e^{-iE_2t}. \label{con-am2}
\end{eqnarray}
As the mixing matrix, $U$ in two generations is $2\times 2$ form and has 4 components, 
it can be parametrized as 
\begin{eqnarray}
U&=&\left(\begin{array}{cc}
e^{i\rho_1} & 0 \\ 
0 & e^{i\rho_2} 
\end{array}\right)
\left(\begin{array}{cc}
\cos \theta & \sin \theta \\ 
-\sin \theta & \cos \theta 
\end{array}\right)\left(\begin{array}{cc}
1 & 0 \\ 
0 & e^{i\phi} 
\end{array}\right)=
\left(\begin{array}{cc}
e^{i\rho_1}\cos \theta & e^{i(\rho_1+\phi)}\sin \theta \\ 
-e^{i\rho_2}\sin \theta & e^{i(\rho_2+\phi)}\cos \theta
\end{array}\right).
\end{eqnarray}
By squaring the amplitudes of (\ref{con-am1}) and (\ref{con-am2}), 
we obtain the well-known form of the oscillation probabilities,  
\begin{eqnarray}
P(\nu_{\alpha L}\to\nu_{\alpha L})&=&|U_{\alpha 1}|^4+|U_{\alpha 2}|^4+2|U_{\alpha 1}U_{\alpha 2}|^2{\rm Re}[e^{i(E_2-E_1)t}] \nonumber \\
&=&c^4+s^4+2s^2c^2\cos (E_2-E_1)t \nonumber \\
&=&1-2s^2c^2[1-\cos(E_2-E_1)t]=1-\sin^2 2\theta \sin^2 \frac{(E_2-E_1)t}{2}, \\
P(\nu_{\alpha L}\to\nu_{\beta L})&=&
|U_{\alpha 1}U_{\beta 1}|^2+|U_{\alpha 2}U_{\beta 2}|^2
+2{\rm Re}[U_{\alpha 1}^*U_{\beta 1}U_{\alpha 2}U_{\beta 2}^*e^{i(E_2-E_1)t}] \nonumber \\
&=&2s^2c^2[1-\cos (E_2-E_1)t]=\sin^2 2\theta\sin^2 \frac{(E_2-E_1)t}{2}, 
\end{eqnarray}
where we use the abbreviation $c=\cos \theta$ and $s=\sin \theta$.
One can see that the phase included in the mixing matrix $U$ does not appear in the oscillation probabilities 
in conventional derivation.
The oscillation probabilities without chirality-flip obtained here are 
interpreted as the transition occurred by the deviation 
between the flavor eigenstates and mass (energy) eigenstates. 

\subsection{Conventional Derivation of Chirality flipped Probabilities}

Next, we review the oscillation probability with chirality-flip calculated by Fukugita-Yanagida 
in one generation \cite{Fukugita-note}.
Fukugita and Yanagida suggested the possibility for the chirality-flip $\nu_L$-$\nu_R$ oscillation in the case that 
the neutrino has finite mass and off-diagonal matrix elements exist in the Hamiltonian.
The Dirac equation in one generation is given by  
\begin{eqnarray}
i\gamma^\mu \partial_\mu \psi-m \psi=0,
\end{eqnarray}
where $m$ is the Dirac mass.
Multiplying $\gamma^0$ from left, the above equation can be rewritten as 
\begin{eqnarray}
&&i\partial_0 \psi+i\gamma^0\gamma^i\partial_i \psi-m \gamma^0\psi=0. 
\end{eqnarray}
Here, if we take the equal momentum assumption, 
$\psi(x,t)=e^{i\vec{p}\cdot \vec{x}}(\nu_{R}^{\prime}, \nu_{R}, \nu_{L}^{\prime}, \nu_{L})^T$ 
and choose as $\overrightarrow{p}=(0,0,p)$,
we obtain the matrix form, 
\begin{eqnarray}
i\frac{d}{dt}
\left(\begin{array}{c}
\nu_{R}^{\prime} \\ \nu_{R} \\ \nu_{L}^{\prime} \\ \nu_{L} 
\end{array}\right)=
\left(\begin{array}{cccc}
p & 0 & m & 0 \\
0 & -p & 0 & m \\
m & 0 & -p & 0 \\
0 & m & 0 & p 
\end{array}\right)\left(\begin{array}{c}
\nu_{R}^{\prime} \\ \nu_{R} \\ \nu_{L}^{\prime} \\ \nu_{L} 
\end{array}\right).
\end{eqnarray}
One can see that the Dirac mass term exist in the off-diagonal components, which  
mixes $\nu_{L}$ and $\nu_{R}$ ($\nu_{L}^{\prime}$ and $\nu_{R}^{\prime}$).
We expect the transition between $\nu_{L}$ and $\nu_{R}$ from the presence of the Dirac mass term.
Extracting the part related to $\nu_{L}$ and $\nu_{R}$, we obtain the equation, 
\begin{eqnarray}
i\frac{d}{dt}\left(\begin{array}{c}
\nu_{R} \\ \nu_{L} 
\end{array}\right) 
=
\left(\begin{array}{cc}
-p & m \\ 
m & p 
\end{array}\right)\left(\begin{array}{c}
\nu_{R} \\ \nu_{L} 
\end{array}\right) \label{LRmix}, 
\end{eqnarray}
and the flavor eigenstates (weak eigenstates) are related to the energy eigenstates by 
the matrix diagonalizing the Hamiltonian in (\ref{LRmix}), 
\begin{eqnarray}
\left(\begin{array}{c}
\nu_{R} \\ \nu_{L} 
\end{array}\right)
=\left(\begin{array}{cc}
\sqrt{\frac{E+p}{2E}} & \sqrt{\frac{E-p}{2E}} \\
-\sqrt{\frac{E-p}{2E}} & \sqrt{\frac{E+p}{2E}} \\
\end{array}\right)\left(\begin{array}{c}
\nu^- \\ \nu^+ 
\end{array}\right), \label{24}
\end{eqnarray}
where 
\begin{eqnarray}
E&=&\sqrt{p^2+m^2}. 
\end{eqnarray}
It is noted that the flavor eigenstates and the mass eigenstates 
are the same in one generation. 
In the case of massless neutrino, the mass eigenstates are coincidence with the energy eigenstates 
completely, but these two kinds of eigenstates are different in the case of massive neutrino.
Therefore, we need to distinguish these eigenstates in future experiments. 
The time development of energy eigenstates are given by 
\begin{eqnarray}
i\frac{d}{dt}\left(\begin{array}{c}
\nu^- \\ \nu^+ 
\end{array}\right)
=\left(\begin{array}{cc}
-E & 0 \\
0 & E \\
\end{array}\right)\left(\begin{array}{c}
\nu^- \\ \nu^+ 
\end{array}\right). 
\end{eqnarray}
Rewriting these relations for fields to those for one particle states, the mass eigenstates after the 
time $t$ are given by  
\begin{eqnarray}
|\nu_{R}(t)\rangle =\sqrt{\frac{E+p}{2E}}e^{iEt} |\nu^-\rangle +\sqrt{\frac{E-p}{2E}}e^{-iEt}|\nu^+\rangle, \\
|\nu_{L}(t)\rangle =-\sqrt{\frac{E-p}{2E}}e^{iEt} |\nu^-\rangle +\sqrt{\frac{E+p}{2E}}e^{-iEt}|\nu^+\rangle, 
\end{eqnarray}
and the conjugate states are also given by, 
\begin{eqnarray}
\langle \nu_{R}| &=&\sqrt{\frac{E+p}{2E}}\langle \nu^-| +\sqrt{\frac{E-p}{2E}}\langle \nu_+|, \\ 
\langle \nu_{L}| &=&-\sqrt{\frac{E-p}{2E}}\langle \nu^-| +\sqrt{\frac{E+p}{2E}}\langle \nu^+|. 
\end{eqnarray}
So, the amplitudes after the time $t$ become 
\begin{eqnarray}
A(\nu_{L}\to\nu_{L})&=&\langle \nu_{L}|\nu_{L}(t)\rangle 
=\frac{E-p}{2E} e^{iEt}+\frac{E+p}{2E}e^{-iEt}, \\
A(\nu_{L}\to\nu_{R})&=&\langle \nu_{R}|\nu_{L}(t)\rangle 
=-\frac{m}{2E} (e^{iEt}-e^{-iEt})=-i\frac{m}{E} \sin (Et).
\end{eqnarray}
Then, the oscillation probabilities are calculated by squaring these amplitudes as  
\begin{eqnarray}
P(\nu_{L}\to\nu_{L})&=&\left(\frac{E-p}{2E}\right)^2 +\left(\frac{E+p}{2E}\right)^2
+\frac{E^2-p^2}{4E^2} (e^{2iEt}+e^{-2iEt}) \nonumber \\
&=&\frac{E^2+p^2}{2E^2}+\frac{m^2}{2E^2}\cos (2Et)=1-\frac{m^2}{2E^2}+\frac{m^2}{2E^2}\cos (2Et) \nonumber \\
&=&1-\frac{m^2}{2E^2}\{1-\cos (2Et)\}=1-\left(\frac{m}{E}\right)^2 \sin^2 (Et),  \\
P(\nu_{L}\to\nu_{R})&=&\left(\frac{m}{E}\right)^2 \sin^2 (Et).
\end{eqnarray}
Fukugita and Yanagida derived these results by using the approximation \cite{Fukugita-note}.
But the results are found to be exact in the above calculation.
About the oscillation between $\nu_L^{\prime}$ and $\nu_R^{\prime}$, 
we can obtain the probabilities by the replacement, $p \to -p$ in (\ref{LRmix}).
The probabilities do not depend on $p$ but $p^2$, therefore the same probabilities are obtained also 
in the case $\nu_{L}^{\prime}$ and $\nu_{R}^{\prime}$. 
Note that the oscillation probabilities with chirality-flip obtained here are 
interpreted as the transition occurred by the deviation 
between the mass (flavor) eigenstates and energy eigenstates.
It is considered that $\nu_{L}\to \nu_{R}$ oscillations hard to happen in the atmospheric neutrino, 
solar neutrino, reactor neutrino, and the accelerator neutrino because the neutrino mass $m$ is 
tiny compared with the energy $E$.
The smallness of the probability is due to the small off-diagonal components in the Hamiltonian 
compared to the difference between diagonal components.
On the other hand, it has been pointed out that the oscillations with chirality-flip have significant 
consequences on the detection of the cosmological relic neutrinos \cite{BBB2020, Ge2020}. 

\section{New Derivation of Oscillation Probabilities}

As reviewed in the previous section, 
the oscillations with and without chirality-flip were calculated separately in previous papers. 
If both types of oscillations exist, considering only one of them will not maintain unitarity.
We can interpret the oscillation without chirality-flip occurred by the deviation between 
the flavor eigenstates and the mass (energy) eigenstates, 
and the oscillation with chirality-flip occurred by the deviation 
between the mass (flavor) eigenstates and the energy eigenstates. 
We would like to consider these oscillations in a unified way in this section.
This can be realized by using the Dirac equation. 
We show that the unitarity holds only when both types of oscillations are considered simultaneously.

\subsection{Oscillation Probabilities of Neutrino}

The lagrangian for two-generation neutrino is represented as 
\begin{eqnarray}
L&=&i\overline{\psi_{\alpha L}}\gamma^\mu \partial_\mu \psi_{\alpha L}+i\overline{\psi_{\alpha R}}
\gamma^\mu \partial_\mu \psi_{\alpha R}
+i\overline{\psi_{\beta L}}\gamma^\mu \partial_\mu \psi_{\beta L}
+i\overline{\psi_{\beta R}}\gamma^\mu \partial_\mu \psi_{\beta R} \nonumber \\
&&-\left[\overline{\psi_{\alpha L}}m_{\alpha\alpha}^*\psi_{\alpha R}+\overline{\psi_{\beta L}}m_{\beta\beta}^*\psi_{\beta R}
+\overline{\psi_{\alpha L}}m_{\beta\alpha}^*\psi_{\beta R}
+\overline{\psi_{\beta L}}m_{\alpha\beta}^*\psi_{\alpha R}\right]\nonumber \\
&&-\left[\overline{\psi_{\alpha R}}m_{\alpha\alpha}\psi_{\alpha L}+\overline{\psi_{\beta R}}m_{\beta\beta}\psi_{\beta L}
+\overline{\psi_{\alpha R}}m_{\alpha\beta}\psi_{\beta L}
+\overline{\psi_{\beta R}}m_{\beta\alpha}\psi_{\alpha L}\right],
\end{eqnarray}
by using 4-component spinors.
The Eular-Lagrange equation for $\overline{\psi_{\alpha L}}$, 
\begin{eqnarray}
\frac{\partial L}{\partial \overline{\psi_{\alpha L}}}
-\partial_\mu \left(\frac{\partial L}{\partial(\partial_\mu \overline{\psi_{\alpha L}})}\right)=0
\end{eqnarray}
gives us a Dirac equation, 
\begin{eqnarray}
i\gamma^\mu \partial_\mu \psi_{\alpha L}-m_{\alpha\alpha}^* \psi_{\alpha R}-m_{\beta\alpha}^* \psi_{\beta R}=0.
\end{eqnarray}
Multiplying $\gamma^0$ from left, we can rewrite the above equation as 
\begin{eqnarray}
&&i\partial_0 \psi_{\alpha L}+i\gamma^0\gamma^i\partial_i \psi_{\alpha L}
-m_{\alpha\alpha}^* \gamma^0\psi_{\alpha R}-m_{\beta\alpha}^* \gamma^0\psi_{\beta R}=0. 
\end{eqnarray}
If we represent this equation by two-components spinors $\xi$ and $\eta$, 
we obtain the following matrix form, 
\begin{eqnarray}
i\partial_0 \left(\begin{array}{c}0 \\ \eta_{\alpha}\end{array}\right)-i
\left(\begin{array}{c}0 \\ \sigma_i\partial_i\eta_{\alpha}\end{array}\right)
-m_{\alpha\alpha}^* \left(\begin{array}{c}0 \\ \xi_{\alpha}\end{array}\right)
-m_{\beta\alpha}^* \left(\begin{array}{c}0 \\ \xi_{\beta}\end{array}\right)=0. 
\end{eqnarray}
We extract the lower part of this equation, 
\begin{eqnarray}
&&i\partial_0 \eta_{\alpha}-i\sigma_i\partial_i \eta_{\alpha}
-m_{\alpha\alpha}^* \xi_{\alpha}-m_{\beta\alpha}^* \xi_{\beta}=0. \label{D.E.of eta}
\end{eqnarray}
In the same way, we also obtain other three equations, 
\begin{eqnarray}
&&i\partial_0 \eta_{\beta}-i\sigma_i\partial_i \eta_{\beta}
-m_{\beta\beta}^* \xi_{\beta}-m_{\alpha\beta}^* \xi_{\alpha}=0, \\
&&i\partial_0 \xi_{\alpha}+i\sigma_i\partial_i \xi_{\alpha}
-m_{\alpha\alpha} \eta_{\alpha}-m_{\alpha\beta} \eta_{\beta}=0, \\
&&i\partial_0 \xi_{\beta}+i\sigma_i\partial_i \xi_{\beta}
-m_{\beta\beta} \eta_{\beta}-m_{\beta\alpha} \eta_{\alpha}=0. \label{D.E.of xi}
\end{eqnarray}
Here, if we take the equal momentum assumption,  
\begin{eqnarray}
\eta_{\alpha}(x,t)&=&e^{i\vec{p}\cdot \vec{x}}\eta_{\alpha}(t)
=e^{i\vec{p}\cdot \vec{x}}\left(\begin{array}{c}\nu_{\alpha L}^{\prime} \\ \nu_{\alpha L}\end{array}\right), \qquad 
\xi_{\alpha}(x,t)=e^{i\vec{p}\cdot \vec{x}}\xi_{\alpha}(t)
=e^{i\vec{p}\cdot \vec{x}}\left(\begin{array}{c}\nu_{\alpha R}^{\prime} \\ \nu_{\alpha R}\end{array}\right), 
\label{etaxia}\\
\eta_{\beta}(x,t)&=&e^{i\vec{p}\cdot \vec{x}}\eta_{\beta}(t)
=e^{i\vec{p}\cdot \vec{x}}\left(\begin{array}{c}\nu_{\beta L}^{\prime} \\ \nu_{\beta L}\end{array}\right), \qquad
\xi_{\beta}(x,t)=e^{i\vec{p}\cdot \vec{x}}\xi_{\beta}(t)
=e^{i\vec{p}\cdot \vec{x}}\left(\begin{array}{c}\nu_{\beta R}^{\prime} \\ \nu_{\beta R}\end{array}\right), 
\label{etaxib}
\end{eqnarray}
and if we choose $\vec{p}=(0,0,p)$, the equations (\ref{D.E.of eta})-(\ref{D.E.of xi}) can be combined into the following matrix form, 
\begin{eqnarray}
i\frac{d}{dt}\left(\begin{array}{c}
\nu_{\alpha R}^{\prime} \\ \nu_{\alpha R} \\ \nu_{\alpha L}^{\prime} \\ \nu_{\alpha L} \\
\nu_{\beta R}^{\prime} \\ \nu_{\beta R} \\ \nu_{\beta L}^{\prime} \\ \nu_{\beta L}
\end{array}\right)
=\left(\begin{array}{cccc|cccc}
p & 0 & m_{\alpha\alpha} & 0 & 0 & 0 & m_{\alpha\beta} & 0 \\ 
0 & -p & 0 & m_{\alpha\alpha} & 0 & 0 & 0 & m_{\alpha\beta} \\
m_{\alpha\alpha}^* & 0 & -p & 0 & m_{\beta\alpha}^* & 0 & 0 & 0 \\ 
0 & m_{\alpha\alpha}^* & 0 & p & 0 & m_{\beta\alpha}^* & 0 & 0 \\
\hline
0 & 0 & m_{\beta\alpha} & 0 & p & 0 & m_{\beta\beta} & 0 \\
0 & 0 & 0 & m_{\beta\alpha} & 0 & -p & 0 & m_{\beta\beta} \\
m_{\alpha\beta}^* & 0 & 0 & 0 & m_{\beta\beta}^* & 0 & -p & 0 \\
0 & m_{\alpha\beta}^* & 0 & 0 & 0 & m_{\beta\beta}^* & 0 & p 
\end{array}\right)\left(\begin{array}{c}
\nu_{\alpha R}^{\prime} \\ \nu_{\alpha R} \\ \nu_{\alpha L}^{\prime} \\ \nu_{\alpha L} \\
\nu_{\beta R}^{\prime} \\ \nu_{\beta R} \\ \nu_{\beta L}^{\prime} \\ \nu_{\beta L}
\end{array}\right). \label{eq50}
\end{eqnarray}
Since the off-diagonal components mixing $\nu_L$ and $\nu_R$ in  (\ref{eq50}) are finite, 
it is expected that the oscillation probability of $\nu_L$ to $\nu_R$ will take a non-zero value. 
We will show it in the following calculation.
On the other hand, the off-diagonal components mixing $\nu_{\alpha L}$ and $\nu_{\beta L}$ 
in the Hamiltonian take a value of zero. Therefore, the oscillations without chirality-flip seem to vanish or be small.
However, contrary to this expectation, the oscillation probabilities are shown to be large.
We can rewrite (\ref{eq50}) by swapping some of the rows and columns as
\begin{eqnarray}
i\frac{d}{dt}\left(\begin{array}{c}
\nu_{\alpha R}^{\prime} \\ \nu_{\beta R}^{\prime} \\ \nu_{\alpha L}^{\prime} \\ \nu_{\beta L}^{\prime} \\
\nu_{\alpha R} \\ \nu_{\beta R} \\ \nu_{\alpha L} \\ \nu_{\beta L}
\end{array}\right)
=\left(\begin{array}{cccc|cccc}
p & 0 & m_{\alpha\alpha} & m_{\alpha\beta} & 0 & 0 & 0 & 0 \\ 
0 & p & m_{\beta\alpha} & m_{\beta\beta} & 0 & 0 & 0 & 0 \\
m_{\alpha\alpha}^* & m_{\beta\alpha}^* & -p & 0 & 0 & 0 & 0 & 0 \\ 
m_{\alpha\beta}^* & m_{\beta\beta}^* & 0 & -p & 0 & 0 & 0 & 0 \\
\hline
0 & 0 & 0 & 0 & -p & 0 & m_{\alpha\alpha} & m_{\alpha\beta} \\
0 & 0 & 0 & 0 & 0 & -p & m_{\beta\alpha} & m_{\beta\beta} \\
0 & 0 & 0 & 0 & m_{\alpha\alpha}^* & m_{\beta\alpha}^* & p & 0 \\
0 & 0 & 0 & 0 & m_{\alpha\beta}^* & m_{\beta\beta}^* & 0 & p 
\end{array}\right)\left(\begin{array}{c}
\nu_{\alpha R}^{\prime} \\ \nu_{\beta R}^{\prime} \\ \nu_{\alpha L}^{\prime} \\ \nu_{\beta L}^{\prime} \\
\nu_{\alpha R} \\ \nu_{\beta R} \\ \nu_{\alpha L} \\ \nu_{\beta L}
\end{array}\right), \label{8-8-nu-matrix}
\end{eqnarray}
where $\nu_{R}$ represents the state with right-handed chirality. 
 In the Standard Model, the flavor of $\nu_R$ cannot be distinguished. 
However, considering the possibility of distinguishable interactions in physics beyond the Standard Model, 
we assign flavor to the right-handed neutrinos as well.
In the above $8\times 8$ matrix, the upper-left $4\times 4$ part is completely separated from the lower-right 
$4\times 4$ part and never mix in this setting even if time passed. 
In the following calculations, we consider the lower-right $4\times 4$ part, 
\begin{eqnarray}
i\frac{d}{dt}\left(\begin{array}{c}
\nu_{\alpha R} \\ \nu_{\beta R} \\ \nu_{\alpha L} \\ \nu_{\beta L}
\end{array}\right)
=\left(\begin{array}{cccc}
-p & 0 & m_{\alpha\alpha} & m_{\alpha\beta} \\
0 & -p & m_{\beta\alpha} & m_{\beta\beta} \\
m_{\alpha\alpha}^* & m_{\beta\alpha}^* & p & 0 \\
m_{\alpha\beta}^* & m_{\beta\beta}^* & 0 & p 
\end{array}\right)\left(\begin{array}{c}
\nu_{\alpha R} \\ \nu_{\beta R} \\ \nu_{\alpha L} \\ \nu_{\beta L}
\end{array}\right),
\end{eqnarray}
where $m_{\alpha\alpha}$, $m_{\beta\beta}$ and $m_{\alpha\beta}$ are complex in general.
The flavor eigenstates are related to the mass eigenstates as
\begin{eqnarray}
\left(\begin{array}{c}
\nu_{\alpha R} \\ \nu_{\beta R} \\ \nu_{\alpha L} \\ \nu_{\beta L}
\end{array}\right)
=\left(\begin{array}{cc|cc}
V_{\alpha 1} & V_{\alpha 2} & 0 & 0 \\
V_{\beta 1} & V_{\beta 2} & 0 & 0 \\
\hline
0 & 0 & U_{\alpha 1} & U_{\alpha 2} \\
0 & 0 & U_{\beta 1} & U_{\beta 2} 
\end{array}\right)\left(\begin{array}{c}
\nu_{1R} \\ \nu_{2R} \\ \nu_{1L} \\ \nu_{2L}
\end{array}\right). \label{flavor-mass}
\end{eqnarray}
(If we cannot distinguish $\nu_{\alpha R}$ with $\nu_{\beta R}$ by any means, 
we can define $\nu_{1R}$ and $\nu_{2R}$ as $\nu_{\alpha R}$ and $\nu_{\beta R}$, 
and then we can regard $V$ as the identity matrix. In this case, the CP phase cannot be 
observed, but we consider the general case here.) 
The Dirac mass term of the Hamiltonian is diagonalized by the above mixing matrix as 
\begin{eqnarray}
\hspace{-0.7cm}
\left(\begin{array}{cc|cc}
V_{\alpha 1}^* & V_{\beta 1}^* & 0 & 0 \\
V_{\alpha 2}^* & V_{\beta 2}^* & 0 & 0 \\
\hline
0 & 0 & U_{\alpha 1}^* & U_{\beta 1}^* \\
0 & 0 & U_{\alpha 2}^* & U_{\beta 2}^* 
\end{array}\right)\!\!\!
\left(\!\!\begin{array}{cccc}
-p & 0 & m_{\alpha\alpha} & m_{\alpha\beta} \\
0 & -p & m_{\beta\alpha} & m_{\beta\beta} \\
m_{\alpha\alpha}^* & m_{\beta\alpha}^* & p & 0 \\
m_{\alpha\beta}^* & m_{\beta\beta}^* & 0 & p 
\end{array}\!\!\right)\!\!\!
\left(\begin{array}{cc|cc}
V_{\alpha 1} & V_{\alpha 2} & 0 & 0 \\
V_{\beta 1} & V_{\beta 2} & 0 & 0 \\
\hline
0 & 0 & U_{\alpha 1} & U_{\alpha 2} \\
0 & 0 & U_{\beta 1} & U_{\beta 2} 
\end{array}\right)\!\!
=\!\!\left(\!\!\begin{array}{cccc}
-p & 0 & m_1 & 0 \\
0 & -p & 0 & m_2 \\
m_1 & 0 & p & 0 \\
0 & m_2 & 0 & p 
\end{array}\!\!\right)\!\!.
\end{eqnarray}
It is emphasized here that the chirality does not change in the transformation (\ref{flavor-mass}). 
In other words, $\nu_{\alpha L}$ and $\nu_{\beta L}$ with left-handed chirality 
can be represented by the linear combination of $\nu_{1L}$ and $\nu_{2L}$ with also left-handed chirality.
This is the reason why we name these two kinds of states ``flavor eigenstates'' and 
``mass eigenstates''. 
The time evolution equation for the mass eigenstates is given by 
\begin{eqnarray}
i\frac{d}{dt}\left(\begin{array}{c}
\nu_{1R} \\ \nu_{2R} \\ \nu_{1L} \\ \nu_{2L}
\end{array}\right)
=\left(\begin{array}{cccc}
-p & 0 & m_1 & 0 \\
0 & -p & 0 & m_2 \\
m_1 & 0 & p & 0 \\
0 & m_2 & 0 & p 
\end{array}\right)\left(\begin{array}{c}
\nu_{1R} \\ \nu_{2R} \\ \nu_{1L} \\ \nu_{2L}
\end{array}\right). \label{eq55}
\end{eqnarray}
Next, let us rewrite the above equation to that for the energy eigenstates 
to diagonalize the Hamiltonian competely.
Exchanging some rows and some columns in (\ref{eq55}), we obtain the equation, 
\begin{eqnarray}
i\frac{d}{dt}\left(\begin{array}{c}
\nu_{1R} \\ \nu_{1L} \\ \nu_{2R} \\ \nu_{2L}
\end{array}\right)
=\left(\begin{array}{cc|cc}
-p & m_1 & 0 & 0 \\
m_1 & p & 0 & 0 \\
\hline  
0 & 0 & -p & m_2 \\
0 & 0 & m_2 & p 
\end{array}\right)\left(\begin{array}{c}
\nu_{1R} \\ \nu_{1L} \\ \nu_{2R} \\ \nu_{2L}
\end{array}\right). 
\end{eqnarray}
We further diagonalize the Hamiltonian in the above equation by replacing the relation 
between the mass eigenstates and the energy eigenstates,  
\begin{eqnarray}
\left(\begin{array}{c}
\nu_{1R} \\ \nu_{1L} \\ \nu_{2R} \\ \nu_{2L}
\end{array}\right)
=\left(\begin{array}{cc|cc}
\sqrt{\frac{E_1+p}{2E_1}} & \sqrt{\frac{E_1-p}{2E_1}} & 0 & 0 \\
-\sqrt{\frac{E_1-p}{2E_1}} & \sqrt{\frac{E_1+p}{2E_1}} & 0 & 0 \\
\hline
0 & 0 & \sqrt{\frac{E_2+p}{2E_2}} & \sqrt{\frac{E_2-p}{2E_2}} \\
0 & 0 & -\sqrt{\frac{E_2-p}{2E_2}} & \sqrt{\frac{E_2+p}{2E_2}} 
\end{array}\right)\left(\begin{array}{c}
\nu_{1}^- \\ \nu_{1}^+ \\ \nu_{2}^- \\ \nu_{2}^+
\end{array}\right), \label{mass-energy}
\end{eqnarray}
and the time evolution equation for the energy eigenstates is given by 
\begin{eqnarray}
i\frac{d}{dt}\left(\begin{array}{c}
\nu_{1}^- \\ \nu_{1}^+ \\ \nu_{2}^- \\ \nu_{2}^+
\end{array}\right)
=\left(\begin{array}{cccc}
-E_1 & 0 & 0 & 0 \\
0 & E_1 & 0 & 0 \\
0 & 0 & -E_2 & 0 \\
0 & 0 & 0 & E_2 
\end{array}\right)\left(\begin{array}{c}
\nu_{1}^- \\ \nu_{1}^+ \\ \nu_{2}^- \\ \nu_{2}^+
\end{array}\right), 
\end{eqnarray}
where 
\begin{eqnarray}
E_1=\sqrt{p^2+m_1^2}, \qquad 
E_2=\sqrt{p^2+m_2^2}. 
\end{eqnarray}
From the equations (\ref{flavor-mass}) and (\ref{mass-energy}), 
the flavor eigenstates are related by the energy eigenstates as 
\begin{eqnarray}
\left(\begin{array}{c}
\nu_{\alpha R} \\ \nu_{\beta R} \\ \nu_{\alpha L} \\ \nu_{\beta L}
\end{array}\right)\!\!\!
&=&\!\!\!
\left(\begin{array}{cc|cc}
V_{\alpha 1} & V_{\alpha 2} & 0 & 0 \\
V_{\beta 1} & V_{\beta 2} & 0 & 0 \\
\hline
0 & 0 & U_{\alpha 1} & U_{\alpha 2} \\
0 & 0 & U_{\beta 1} & U_{\beta 2} 
\end{array}\right)\!\!\!
\left(\begin{array}{cccc}
1 & 0 & 0 & 0 \\
0 & 0 & 1 & 0 \\
0 & 1 & 0 & 0 \\
0 & 0 & 0 & 1 
\end{array}\right)\!\!\!
\left(\begin{array}{cc|cc}
\sqrt{\frac{E_1+p}{2E_1}} & \sqrt{\frac{E_1-p}{2E_1}} & 0 & 0 \\
-\sqrt{\frac{E_1-p}{2E_1}} & \sqrt{\frac{E_1+p}{2E_1}} & 0 & 0 \\
\hline
0 & 0 & \sqrt{\frac{E_2+p}{2E_2}} & \sqrt{\frac{E_2-p}{2E_2}} \\
0 & 0 & -\sqrt{\frac{E_2-p}{2E_2}} & \sqrt{\frac{E_2+p}{2E_2}} 
\end{array}\right)\!\!\!\left(\begin{array}{c}
\nu_{1}^- \\ \nu_{1}^+ \\ \nu_{2}^- \\ \nu_{2}^+
\end{array}\right) \nonumber \\
&=&\left(\begin{array}{cc|cc}
\sqrt{\frac{E_1+p}{2E_1}}V_{\alpha 1} & \sqrt{\frac{E_1-p}{2E_1}}V_{\alpha 1} 
& \sqrt{\frac{E_2+p}{2E_2}}V_{\alpha 2} & \sqrt{\frac{E_2-p}{2E_2}}V_{\alpha 2} \\
\sqrt{\frac{E_1+p}{2E_1}}V_{\beta 1} & \sqrt{\frac{E_1-p}{2E_1}}V_{\beta 1} 
& \sqrt{\frac{E_2+p}{2E_2}}V_{\beta 2} & \sqrt{\frac{E_2-p}{2E_2}}V_{\beta 2} \\
\hline
-\sqrt{\frac{E_1-p}{2E_1}}U_{\alpha 1} & \sqrt{\frac{E_1+p}{2E_1}}U_{\alpha 1} 
& -\sqrt{\frac{E_2-p}{2E_2}}U_{\alpha 2} & \sqrt{\frac{E_2+p}{2E_2}}U_{\alpha 2} \\
-\sqrt{\frac{E_1-p}{2E_1}}U_{\beta 1} & \sqrt{\frac{E_1+p}{2E_1}}U_{\beta 1} 
& -\sqrt{\frac{E_2-p}{2E_2}}U_{\beta 2} & \sqrt{\frac{E_2+p}{2E_2}}U_{\beta 2} 
\end{array}\right)
\left(\begin{array}{c}
\nu_{1}^- \\ \nu_{1}^+ \\ \nu_{2}^- \\ \nu_{2}^+
\end{array}\right). 
\end{eqnarray}
Rewriting these relations for fields to those for one particle states, 
the flavor eigenstates after the time $t$ are given by 
\begin{eqnarray}
|\nu_{\alpha R}(t)\rangle &=&V_{\alpha 1}^*\sqrt{\frac{E_1+p}{2E_1}}e^{iE_1t}|\nu_1^-\rangle 
+V_{\alpha 1}^*\sqrt{\frac{E_1-p}{2E_1}}e^{-iE_1t} |\nu_1^+\rangle \nonumber \\
&&+V_{\alpha 2}^*\sqrt{\frac{E_2+p}{2E_2}}e^{iE_2t}|\nu_2^-\rangle 
+V_{\alpha 2}^*\sqrt{\frac{E_2-p}{2E_2}}e^{-iE_2t}|\nu_2^+\rangle, \\
|\nu_{\beta R}(t)\rangle &=&V_{\beta 1}^*\sqrt{\frac{E_1+p}{2E_1}}e^{iE_1t}|\nu_1^-\rangle 
+V_{\beta 1}^*\sqrt{\frac{E_1-p}{2E_1}}e^{-iE_1t}|\nu_1^+\rangle \nonumber \\  
&&+V_{\beta 2}^*\sqrt{\frac{E_2+p}{2E_2}}e^{iE_2t}|\nu_2^-\rangle 
+V_{\beta 2}^*\sqrt{\frac{E_2-p}{2E_2}}e^{-iE_2t}|\nu_2^+\rangle,  \\
|\nu_{\alpha L}(t)\rangle &=&-U_{\alpha 1}^*\sqrt{\frac{E_1-p}{2E_1}}e^{iE_1t}|\nu_1^-\rangle 
+U_{\alpha 1}^*\sqrt{\frac{E_1+p}{2E_1}}e^{-iE_1t}|\nu_1^+\rangle \nonumber \\  
&&-U_{\alpha 2}^*\sqrt{\frac{E_2-p}{2E_2}}e^{iE_2t}|\nu_2^-\rangle 
+U_{\alpha 2}^*\sqrt{\frac{E_2+p}{2E_2}}e^{-iE_2t}|\nu_2^+\rangle,  \\
|\nu_{\beta L}(t)\rangle &=&-U_{\beta 1}^*\sqrt{\frac{E_1-p}{2E_1}}e^{iE_1t}|\nu_1^-\rangle 
+U_{\beta 1}^*\sqrt{\frac{E_1+p}{2E_1}}e^{-iE_1t}|\nu_1^+\rangle \nonumber \\  
&&-U_{\beta 2}^*\sqrt{\frac{E_2-p}{2E_2}}e^{iE_2t}|\nu_2^-\rangle 
+U_{\beta 2}^*\sqrt{\frac{E_2+p}{2E_2}}e^{-iE_2t}|\nu_2^+\rangle,  
\end{eqnarray}
and we also have the conjugate states,  
\begin{eqnarray}
\hspace{-0.5cm}\langle \nu_{\alpha R}| &=&V_{\alpha 1}\sqrt{\frac{E_1+p}{2E_1}}\langle \nu_1^-| 
+V_{\alpha 1}\sqrt{\frac{E_1-p}{2E_1}}\langle \nu_1^+|
+V_{\alpha 2}\sqrt{\frac{E_2+p}{2E_2}}\langle \nu_2^-|
+V_{\alpha 2}\sqrt{\frac{E_2-p}{2E_2}}\langle \nu_2^+|, \\
\hspace{-0.5cm}\langle \nu_{\beta R}| &=&V_{\beta 1}\sqrt{\frac{E_1+p}{2E_1}}\langle \nu_1^-| 
+V_{\beta 1}\sqrt{\frac{E_1-p}{2E_1}}\langle \nu_1^+|
+V_{\beta 2}\sqrt{\frac{E_2+p}{2E_2}}\langle \nu_2^-| 
+V_{\beta 2}\sqrt{\frac{E_2-p}{2E_2}}\langle \nu_2^+|,  \\
\hspace{-0.5cm}\langle \nu_{\alpha L}| &=&-U_{\alpha 1}\sqrt{\frac{E_1-p}{2E_1}}\langle \nu_1^-| 
+U_{\alpha 1}\sqrt{\frac{E_1+p}{2E_1}}\langle \nu_1^+|
-U_{\alpha 2}\sqrt{\frac{E_2-p}{2E_2}}\langle \nu_2^-| 
+U_{\alpha 2}\sqrt{\frac{E_2+p}{2E_2}}\langle \nu_2^+|,  \\
\hspace{-0.5cm}\langle \nu_{\beta L}| &=&-U_{\beta 1}\sqrt{\frac{E_1-p}{2E_1}}\langle \nu_1^-| 
+U_{\beta 1}\sqrt{\frac{E_1+p}{2E_1}}\langle \nu_1^+|
-U_{\beta 2}\sqrt{\frac{E_2-p}{2E_2}}\langle \nu_2^-| 
+U_{\beta 2}\sqrt{\frac{E_2+p}{2E_2}}\langle \nu_2^+|.  
\end{eqnarray}
It is crucial that the flavor eigenstates are constituted by not only positive energy states 
but also negative energy states. 
This is because the positive energy states and the negative energy states are included in the same multiplet. 
Then, the amplitudes are calculated by 
\begin{eqnarray}
A(\nu_{\alpha L}\to\nu_{\alpha L})&=&\langle \nu_{\alpha L}|\nu_{\alpha L}(t)\rangle \nonumber \\
&&\hspace{-2.8cm}=|U_{\alpha 1}|^2\frac{E_1-p}{2E_1}e^{iE_1t}
+|U_{\alpha 1}|^2\frac{E_1+p}{2E_1}e^{-iE_1t}
+|U_{\alpha 2}|^2\frac{E_2-p}{2E_2}e^{iE_2t}
+|U_{\alpha 2}|^2\frac{E_2+p}{2E_2}e^{-iE_2t} \nonumber \\
&&\hspace{-2.8cm}=|U_{\alpha 1}|^2\cos (E_1t)-i|U_{\alpha 1}|^2\cdot \frac{p}{E_1}\sin (E_1t)
+|U_{\alpha 2}|^2\cos (E_2t)-i|U_{\alpha 2}|^2\cdot \frac{p}{E_2}\sin (E_2t), \\
A(\nu_{\alpha L}\to\nu_{\beta L})&=&\langle \nu_{\beta L}|\nu_{\alpha L}(t)\rangle \nonumber \\
&&\hspace{-2.8cm}=U_{\alpha 1}^*U_{\beta 1}\left(\frac{E_1-p}{2E_1}e^{iE_1t}+\frac{E_1+p}{2E_1}e^{-iE_1t}\right)
+U_{\alpha 2}^*U_{\beta 2}\left(\frac{E_2-p}{2E_2}e^{iE_2t}+\frac{E_2+p}{2E_2}e^{-iE_2t}\right) \nonumber \\
&&\hspace{-2.8cm}=U_{\alpha 1}^*U_{\beta 1}\left\{\cos (E_1t)-i\frac{p}{E_1}\sin (E_1t)\right\}
+U_{\alpha 2}^*U_{\beta 2}\left\{\cos (E_2t)-i\frac{p}{E_2}\sin (E_2t)\right\}, \\
A(\nu_{\alpha L}\to\nu_{\alpha R})&=&\langle \nu_{\alpha R}|\nu_{\alpha L}(t)\rangle \nonumber \\
&&\hspace{-2.8cm}=-U_{\alpha 1}^*V_{\alpha 1}\frac{m_1}{2E_1}(e^{iE_1t}-e^{-iE_1t})
-U_{\alpha 2}^*V_{\alpha 2}\frac{m_2}{2E_2}(e^{iE_2t}-e^{-iE_2t}) \nonumber \\
&&\hspace{-2.8cm}=-iU_{\alpha 1}^*V_{\alpha 1}\frac{m_1}{E_1}\sin (E_1t)
-iU_{\alpha 2}^*V_{\alpha 2}\frac{m_2}{E_2}\sin (E_2t), \\
A(\nu_{\alpha L}\to\nu_{\beta R})&=&\langle \nu_{\beta R}|\nu_{\alpha L}(t)\rangle \nonumber \\
&&\hspace{-2.8cm}=U_{\alpha 1}^*V_{\beta 1}\frac{m_1}{E_1}(e^{iE_1t}-e^{-iE_1t})
-U_{\alpha 2}^*V_{\beta 2}\frac{m_2}{E_2}(e^{iE_2t}-e^{-iE_2t})\nonumber \\
&&\hspace{-2.8cm}=-iU_{\alpha 1}^*V_{\beta 1}\frac{m_1}{E_1}\sin (E_1t)
-iU_{\alpha 2}^*V_{\beta 2}\frac{m_2}{E_2}\sin (E_2t). 
\end{eqnarray}
Furthermore, we obtain the oscillation probabilies by squaring these amplitudes, 
\begin{eqnarray}
&&\hspace{-0.8cm}P(\nu_{\alpha L}\to\nu_{\alpha L})
=\{|U_{\alpha 1}|^2\cos (E_1t)+|U_{\alpha 2}|^2\cos (E_2t)\}^2 \nonumber \\
&&\hspace{-0.3cm}+\left\{|U_{\alpha 1}|^2\cdot \frac{p}{E_1}\sin (E_1t)
+|U_{\alpha 2}|^2\cdot \frac{p}{E_2}\sin (E_2t)\right\}^2, \label{eq82}\\
&&\hspace{-0.8cm}P(\nu_{\alpha L}\to\nu_{\beta L})=
|U_{\alpha 1}U_{\beta 1}|^2\left\{\cos^2 (E_1t)+\frac{p^2}{E_1^2}\sin^2 (E_1t)\right\} \nonumber \\
&&\hspace{-0.3cm}+|U_{\alpha 2}U_{\beta 2}|^2\left\{\cos^2 (E_2t)+\frac{p^2}{E_2^2}\sin^2 (E_2t)\right\} \nonumber \\
&&\hspace{-0.3cm}+2{\rm Re}\left[U_{\alpha 1}^*U_{\beta 1}\left\{\cos (E_1t)-i\frac{p}{E_1}\sin (E_1t)\right\}
U_{\alpha 2}U_{\beta 2}^*\left\{\cos (E_2t)+i\frac{p}{E_2}\sin (E_2t)\right\}\right], \label{eq83} 
\end{eqnarray}
\begin{eqnarray}
&&\hspace{-0.8cm}P(\nu_{\alpha L}\to\nu_{\alpha R})=
|U_{\alpha 1}V_{\alpha 1}|^2\frac{m_1^2}{E_1^2}\sin^2 (E_1t)+|U_{\alpha 2}V_{\alpha 2}|^2\frac{m_2^2}{E_2^2}\sin^2 (E_2t) \nonumber \\
&&\hspace{-0.3cm}+2{\rm Re}[U_{\alpha 1}U_{\alpha 2}^*V_{\alpha 1}^*V_{\alpha 2}]\frac{m_1m_2}{E_1E_2}\sin (E_1t)\sin (E_2t), \label{eq84}\\
&&\hspace{-0.8cm}P(\nu_{\alpha L}\to\nu_{\beta R})=
|U_{\alpha 1}V_{\beta 1}|^2\frac{m_1^2}{E_1^2}\sin^2 (E_1t)
+|U_{\alpha 2}V_{\beta 2}|^2\frac{m_2^2}{E_2^2}\sin^2 (E_2t) \nonumber \\
&&\hspace{-0.3cm}+2{\rm Re}[U_{\alpha 1}U_{\alpha 2}^*V_{\beta 1}^*V_{\beta 2}]\frac{m_1m_2}{E_1E_2}\sin (E_1t)\sin (E_2t). \label{eq85}
\end{eqnarray}
In general, $2\times 2$ unitary matrices $U$ and $V$ are parametrized by 
\begin{eqnarray}
\hspace{-0.5cm}U\!\!&=&\!\!\left(\!\!\begin{array}{cc}
e^{i\rho_{1L}} & 0 \\ 
0 & e^{i\rho_{2L}} 
\end{array}\!\!\right)\!\!
\left(\begin{array}{cc}
\cos \theta_L & \sin \theta_L \\ 
-\sin \theta_L & \cos \theta_L 
\end{array}\right)\!\!\left(\begin{array}{cc}
1 & 0 \\ 
0 & e^{i\phi_L} 
\end{array}\!\!\right)
\!\!=\!\!
\left(\begin{array}{cc}
e^{i\rho_{1L}}\cos \theta_L & e^{i(\rho_{1L}+\phi_L)}\sin \theta_L \\ 
-e^{i\rho_{2L}}\sin \theta_L & e^{i(\rho_{2L}+\phi_L)}\cos \theta_L 
\end{array}\!\!\right),
\\
\hspace{-0.5cm}V\!\!&=&\!\!\left(\!\!\begin{array}{cc}
e^{i\rho_{1R}} & 0 \\ 
0 & e^{i\rho_{2R}} 
\end{array}\!\!\right)\!\!
\left(\begin{array}{cc}
\cos \theta_R & \sin \theta_R \\ 
-\sin \theta_R & \cos \theta_R 
\end{array}\right)\!\!\left(\begin{array}{cc}
1 & 0 \\ 
0 & e^{i\phi_R} 
\end{array}\!\!\right)\!\!
=\!\!
\left(\!\!\begin{array}{cc}
e^{i\rho_{1R}}\cos \theta_R & e^{i(\rho_{1R}+\phi_R)}\sin \theta_R \\ 
-e^{i\rho_{2R}}\sin \theta_R & e^{i(\rho_{2R}+\phi_R)}\cos \theta_R 
\end{array}\!\!\right),
\end{eqnarray}
respectively.
Substituting these representations into (\ref{eq82})-(\ref{eq85}), the oscillation probabilities for 
$\nu_{\alpha L}$ to $\nu_{\alpha L}$, $\nu_{\beta L}$, $\nu_{\alpha R}$ and $\nu_{\beta R}$ 
are given by 
\begin{eqnarray}
&&\hspace{-0.5cm}P(\nu_{\alpha L}\to\nu_{\alpha L})=
1-4s_L^2c_L^2\sin^2 \frac{(E_2-E_1)t}{2} \label{old-survive} \\
&&\hspace{-0.5cm}-\left[c_L^4\cdot \frac{m_1^2}{E_1^2}\sin^2 (E_1t)
+s_L^4\cdot \frac{m_2^2}{E_2^2}\sin^2 (E_2t)+2s_L^2c_L^2 \left(\!1-\frac{p^2}{E_1E_2}\right)
\sin (E_1t)\sin (E_2t)\right]\!, \label{new-survive} \\
&&\hspace{-0.5cm}P(\nu_{\alpha L}\to\nu_{\beta L})=
4s_L^2c_L^2\sin^2 \frac{(E_2-E_1)t}{2} \label{old-transition} \\
&&\hspace{0cm}-s_L^2c_L^2\left[\frac{m_1^2}{E_1^2}\sin^2 (E_1t)+\frac{m_2^2}{E_2^2}\sin^2 (E_2t)
-2\left(1-\frac{p^2}{E_1E_2}\right)\sin (E_1t)\sin (E_2t)\right], \label{new-transition} \\
&&\hspace{-0.5cm}P(\nu_{\alpha L}\to\nu_{\alpha R})=
c_L^2c_R^2\frac{m_1^2}{E_1^2}\sin^2 (E_1t)+s_L^2s_R^2\frac{m_2^2}{E_2^2}\sin^2 (E_2t) \nonumber \\
&&\hspace{2cm}+2s_Ls_Rc_Lc_R\cos (\phi_R-\phi_L)\frac{m_1m_2}{E_1E_2}\sin (E_1t)\sin (E_2t), \label{h-change-survive} \\
&&\hspace{-0.5cm}P(\nu_{\alpha L}\to\nu_{\beta R})=
c_L^2s_R^2\frac{m_1^2}{E_1^2}\sin^2 (E_1t)
+s_L^2c_R^2\frac{m_2^2}{E_2^2}\sin^2 (E_2t) \nonumber \\
&&\hspace{2cm}-2s_Ls_Rc_Lc_R\cos (\phi_R-\phi_L)\frac{m_1m_2}{E_1E_2}\sin (E_1t)\sin (E_2t), \label{h-change-transition}
\end{eqnarray}
where we use the abbreviation $\sin \theta_L=s_L$ and $\cos \theta_L=c_L$ and so on. 
As shown above, we derived the oscillation probabilities for both the case with chirality-flip 
and the case without chirality-flip in a unified way by using the Dirac equation.
The oscillation probabilities obtained above are exact ones based on the equal momentum assumption.
Taking the limit $m/E\to 0$, one can see that these oscillation probabilities coincide with the well-known probabilities 
derived by the Schr$\ddot{\rm o}$dinger equation.  

Here, we list some crucial points in order.
First, in the probabilities without chirality-flip, the new terms (\ref{new-survive}) and (\ref{new-transition}) 
appear in addition to the well-known terms (\ref{old-survive}) and (\ref{old-transition}).
At the same time, the probabilities with chirality-flip are in order $O(m^2/E^2)$. 
As shown later, these terms in order $O(m^2/E^2)$ cancel out each other and 
the sum of all probabilities is kept in one.
Second, we found that the oscillation probabilities depend on not only the mass squared differences 
but also the absolute value of mass in exact expressions.
Third, the new CP phase appears in the probabilities even for Dirac neutrino in two generations if there exist some new interactions which distinguish the flavor of right-handed neutrinos. 
This new phase appears in the probabilities with chirality-flip and can be interpreted to be a comprehensive phase including the Majorana phase.

We mention the differences between previous studies and our results. The $P(\nu_{\alpha L}\to \nu_{\beta L})$ was calculated relativistically in \cite{Dvornikov} up to order $O(m^2/E^2)$. Furthermore, in \cite{Bernardini2004, Bernardini2005, Nishi}, 
$P(\nu_{\alpha L} \to \nu_{\beta L})$ was derived without any approximations using the wave-packet formalism. 
After some calculations, it was confirmed that their results are consistent with our expressions (\ref{old-transition}) and (\ref{new-transition}). In addition to $P(\nu_{\alpha L}\to \nu_{\beta L})$, we also calculated $P(\nu_{\alpha L} \to \nu_{\beta R})$, discovering that the latter depends on the new CP phase and the mixing angle of right-handed neutrinos.

Let us comment about $\nu_L^{\prime}$ and $\nu_R^{\prime}$ in the end of this subsection.
In order to obtain the probabilities for $\nu^{\prime}$, we only have to replace $p$ to $-p$ 
in (\ref{old-survive})-(\ref{h-change-transition}).
As the oscillation probabilities only depend on $p^2$, we obtain the same probabilities as those for 
$\nu_L$ and $\nu_R$ in this framework.

\subsection{Energy and Baseline Where New Terms give Non-trivial Contribution}

From the results obtained above, it is very interesting to explore the case that the new terms have a non-trivial contribution. 
We discuss under what situations the contribution of the new terms is of comparable magnitude to that of the usual terms.
Replacing the time $t$ to the distance $L$ that neutrinos travel in the time $t$, 
the leading terms contributing to usual neutrino oscillations have the magnitude of order  
$\sin^2 (m^2 L/E)$. On the other hand, the new terms have the contribution with the magnitude of order  
$m^2/E^2\sin^2 (EL)$.
Below, let us divide into three cases by the value of $EL$ and $m^2L/E$, 
and compare the magnitude of the two kinds of terms.

\begin{enumerate}
\item[(i)] Case for $EL<1$, i.e. case for $\displaystyle{L<10^{-6}{\rm m}\cdot \frac{\rm eV}{E}}$\\
If the contribution of the new terms and the usual terms are nearly equal, we have the relation, 
$\displaystyle{\sin^2 \left(\frac{m^2L}{E}\right)\simeq \frac{m^2}{E^2}\sin^2 (EL)}$ and 
that is approximated by \\ 
$\displaystyle{\left(\frac{m^2L}{E}\right)^2\simeq \frac{m^2}{E^2}(EL)^2}$ 
$\Longleftrightarrow$ $\displaystyle{\frac{m^2}{E^2}\simeq 1}$. \\
This means that new terms become important for the case that the mass of neutrino is comparative with the energy.
\item[(ii)] Case for $EL>1$ and $\displaystyle{\frac{m^2L}{E}}<1$, i.e. 
case for $\displaystyle{10^{-6}{\rm m}\cdot \frac{\rm eV}{E}<L<10^{-6}{\rm m}\cdot 
\frac{E}{\rm eV}\cdot \frac{{\rm eV}^2}{m^2}}$ \\
The relation, $\displaystyle{\sin^2 \left(\frac{m^2L}{E}\right)<\frac{m^2}{E^2}\sin^2 (EL)}$ is approximated by 
$\displaystyle{\left(\frac{m^2L}{E}\right)^2< \frac{m^2}{E^2}}$ 
$\Longleftrightarrow$ $\displaystyle{mL< 1}$ 
$\Longleftrightarrow$ $\displaystyle{m({\rm eV}^2)\cdot L({\rm m})<10^{-6}}$. 
If we set the typical neutrino mass about $m^2\simeq 10^{-3}({\rm eV}^2)$, the contribution of the new terms 
is important in the range $\displaystyle{L({\rm m})<10^{-3}}$ almost without depending on energy.
\item[(iii)] Case for $EL>1$ and $\displaystyle{\frac{m^2L}{E}}>1$, i.e.
case for $\displaystyle{L>10^{-6}{\rm m}\cdot \frac{\rm eV}{E}}$ and 
$\displaystyle{L>10^{-6}{\rm m}\cdot \frac{E}{\rm eV}\cdot \frac{{\rm eV}^2}{m^2}}$\\
The relation, $\displaystyle{\sin^2 \left(\frac{m^2L}{E}\right)\simeq \frac{m^2}{E^2}\sin^2 (EL)}$ is approximated by 
$\displaystyle{\frac{m^2}{E^2}\simeq 1}$. 

We found that the new terms are important for the case where the neutrino mass is of the same order as the energy, as in (i).
However, if the distance $L$ becomes too long the wave packets of light neutrinos and heavy neutrinos separate, and the interference, i.e. neutrino oscillations do not occur. 
\end{enumerate}

In previous neutrino experiments, the baseline length has been comparatively large, and the neutrino mass is small enough to be ignored compared to the neutrino energy, which allows us to neglect the contribution of the new terms.
However, we may observe the contribution of the new terms and obtain the information of the absolute mass of 
neutrino if we measure the neutrino oscillation probabilities 
in atomic size like $0\nu\beta\beta$ decay experiments.
Besides, the contribution of the new CP phase and the mixing angle may also appear 
if some interactions which distinguish the flavor of 
right-handed neutrinos exist.
An example, in which $\nu_R$ has weak interaction as $\nu_L$, is the case of Majorana neutrinos.  
In this case, $\nu_R$ is identified with $\nu_L^c$ and therefore, the mixing and the CP phase of 
$\nu_R$ relate to those of $\nu_L$ like $\phi_R=-\phi_L$.
See our next paper for detail.
At the end of this subsection, let us comment that the left-right symmetric models were considered 
in the framework of the Dirac neutrinos \cite{Borah2017} 
and there is a possibility for the mixing angle and the CP phase of $\nu_R$ to be 
independent of $\nu_L$. 

\subsection{Unitarity Check of Oscillation Probabilities}

In the Standard Model, $\nu_R$ can be identified with the mass eigenstate as it does not have weak interactions.
In this situation, $V$ becomes the identity matrix and we can choose $s_R=0$. 
Therefore, the mixing angle for $\nu_R$ and the CP phase do not appear and then 
only the sum of two oscillation probabilities, 
\begin{eqnarray}
\hspace{-0.7cm}
P(\nu_{\alpha L}\to\nu_{R})=P(\nu_{\alpha L}\to\nu_{\alpha R})+P(\nu_{\alpha L}\to\nu_{\beta R})=
c_L^2\frac{m_1^2}{E_1^2}\sin^2 (E_1t)+s_L^2\frac{m_2^2}{E_2^2}\sin^2 (E_2t)
\end{eqnarray}
can be observable. Similarly, summing the probabilities of oscillating to $\nu_{\alpha L}$ and $\nu_{\beta L}$, we obtain 
\begin{eqnarray}
P(\nu_{\alpha L}\to\nu_{L})&=&P(\nu_{\alpha L}\to\nu_{\alpha L})+P(\nu_{\alpha L}\to\nu_{\beta L}) \nonumber \\
&&\hspace{-2.5cm}=
\left\{c_L^2\cos^2 (E_1t)+s_L^2\cos^2 (E_2t)\right\}+\left\{c_L^2\cdot \frac{p^2}{E_1^2}\sin^2 (E_1t)
+s_L^2\cdot \frac{p^2}{E_2^2}\sin^2 (E_2t)\right\}.
\end{eqnarray}
Furthermore, the total sum of the oscillation probabilities from $\nu_{\alpha L}$ is calculated as  
\begin{eqnarray}
P(\nu_{\alpha L}\to\nu_{R})+P(\nu_{\alpha L}\to\nu_{L})
&=&c_L^2\frac{m_1^2}{E_1^2}\sin^2 (E_1t)
+s_L^2\frac{m_2^2}{E_2^2}\sin^2 (E_2t) \nonumber \\
&&\hspace{-5cm}+\left\{c_L^2\cos^2 (E_1t)+s_L^2\cos^2 (E_2t)\right\} 
+\left\{c_L^2\cdot \frac{E_1^2-m_1^2}{E_1^2}\sin^2 (E_1t)
+s_L^2\cdot \frac{E_2^2-m_2^2}{E_2^2}\sin^2 (E_2t)\right\}\nonumber \\
&&\hspace{-5cm}=
\left\{c_L^2\cos^2 (E_1t)+s_L^2\cos^2 (E_2t)\right\}+\left\{c_L^2\sin^2 (E_1t)+s_L^2\sin^2 (E_2t)\right\}
=c_L^2+s_L^2=1.
\end{eqnarray}
As expected, we have confirmed that the total sum of the probabilities with and without chirality-flip is exactly one.

\subsection{Oscillation Probabilities of Right-Handed Neutrinos}
 
If the flavor of right-handed neutrinos can be measured by some methods in physics beyond the Standard Model, $\nu_R$ oscillations would be observed. The probabilities of these oscillations can be calculated in the same way as for left-handed neutrinos. 
For completeness, we write down these probabilities.
Replacing $L$ to $R$ in the probabilities of $\nu_L$, we have the following expressions, 
\begin{eqnarray}
&&\hspace{-1cm}P(\nu_{\alpha R}\to\nu_{\alpha R})
=
1-4s_R^2c_R^2\sin^2 \frac{(E_2-E_1)t}{2} \nonumber \\
&&\hspace{-1cm}-\left[c_R^4\cdot \frac{m_1^2}{E_1^2}\sin^2 (E_1t)
+s_R^4\cdot \frac{m_2^2}{E_2^2}\sin^2 (E_2t)+2s_R^2c_R^2\left(1-\frac{p^2}{E_1E_2}\right)
\sin (E_1t)\sin (E_2t)\right], \\
&&\hspace{-1cm}P(\nu_{\alpha R}\to\nu_{\beta R})
=4s_R^2c_R^2\sin^2 \frac{(E_2-E_1)t}{2}\nonumber \\
&&-s_R^2c_R^2\left[\frac{m_1^2}{E_1^2}\sin^2 (E_1t)+\frac{m_2^2}{E_2^2}\sin^2 (E_2t)
-2\left(1-\frac{p^2}{E_1E_2}\right)\sin (E_1t)\sin (E_2t)\right],\\
&&\hspace{-1cm}P(\nu_{\alpha R}\to\nu_{\alpha L})=
c_L^2c_R^2\frac{m_1^2}{E_1^2}\sin^2 (E_1t)+s_L^2s_R^2\frac{m_2^2}{E_2^2}\sin^2 (E_2t) \nonumber \\
&&+2s_Ls_Rc_Lc_R\cos (\phi_R-\phi_L)\frac{m_1m_2}{E_1E_2}\sin (E_1t)\sin (E_2t), \\
&&\hspace{-1cm}P(\nu_{\alpha R}\to\nu_{\beta L})=
c_R^2s_L^2\frac{m_1^2}{E_1^2}\sin^2 (E_1t)
+s_R^2c_L^2\frac{m_2^2}{E_2^2}\sin^2 (E_2t) \nonumber \\
&&-2s_Ls_Rc_Lc_R\cos (\phi_R-\phi_L)\frac{m_1m_2}{E_1E_2}\sin (E_1t)\sin (E_2t).
\end{eqnarray}
The new mixing angle of the right-handed neutrinos can be observed through $\nu_R \to \nu_R$ oscillations 
if $W_R$ exists in the left-right symmetric model.
In this case, one may produce $\nu_R$ beam by decaying $W_R$ generated from the collision of 
the high energy pi-mesons to the target. 
If the $\nu_R$ reacts the matter in the detector and reproduces the charged lepton with right-handed through $W_R$, 
we have a chance to measure this new mixing angle without suppression due to small neutrino masses.

\subsection{Oscillation Probabilities of Anti-Neutrinos}

Next, we summarize the oscillation probabilities of anti-neutrinos. 
Charge conjugation of $\psi$, 
\begin{eqnarray}
\psi^c=(\psi_L)^c+(\psi_R)^c
\equiv \left(\begin{array}{c}\nu_L^c \\ \nu_L^{c\prime} \\ \nu_R^c \\ \nu_R^{c\prime} \end{array}\right)
\equiv \left(\begin{array}{c}i\sigma_2 \eta^* \\ -i\sigma_2 \xi^{*} \end{array}\right)
=\left(\begin{array}{c}\nu_L^* \\ -\nu_L^{*\prime} \\ -\nu_R^* \\ \nu_R^{*\prime} \end{array}\right),
\end{eqnarray}
also satisfies a Dirac equation, 
\begin{eqnarray}
i\gamma^\mu \partial_\mu \psi_{\alpha L}^c-m_{\alpha\alpha} \psi_{\alpha R}^c-m_{\beta\alpha} \psi_{\beta R}^c=0.
\end{eqnarray}
It is noted that the Dirac equation satisfied by the charge conjugated field $\psi^c$ is a little bit different from 
that for the original field $\psi$ because the mass terms are complex in general.
Namely, the Dirac equation for $\psi^c$ has the complex conjugate mass terms against the original equation.
Multiplying $\gamma^0$ from the left, we obtain 
\begin{eqnarray}
&&i\partial_0 \psi_{\alpha L}^c+i\gamma^0\gamma^i\partial_i \psi_{\alpha L}^c
-m_{\alpha\alpha} \gamma^0\psi_{\alpha R}^c-m_{\beta\alpha} \gamma^0\psi_{\beta R}^c=0. 
\end{eqnarray}
If we represent this equation by two-components spinors $\xi$ and $\eta$, we obtain the following matrix form, 
\begin{eqnarray}
i\partial_0 \left(\!\!\begin{array}{c}i\sigma_2\eta_{\alpha}^* \\ 0\end{array}\!\!\right)
+i\left(\!\!\begin{array}{c}\sigma_i\partial_i (i\sigma_2\eta_{\alpha}^*) \\ 0\end{array}\!\!\right)
-m_{\alpha\alpha} \left(\begin{array}{c}-i\sigma_2\xi_{\alpha}^* \\ 0\end{array}\right)
-m_{\beta\alpha} \left(\begin{array}{c}-i\sigma_2\xi_{\beta}^* \\ 0\end{array}\right)=0, 
\end{eqnarray}
and extracting the upper part, we obtain 
\begin{eqnarray}
i\partial_0 (i\sigma_2\eta_{\alpha}^*)+i\sigma_i\partial_i (i\sigma_2\eta_{\alpha}^*)
-m_{\alpha\alpha} (-i\sigma_2\xi_{\alpha}^*)-m_{\beta\alpha} (-i\sigma_2\xi_{\beta}^*)=0. \label{D.E.anti-nu1}
\end{eqnarray}
In the same way, we also obtain three equations, 
\begin{eqnarray}
&&i\partial_0 (i\sigma_2\eta_{\beta}^*)+i\sigma_i\partial_i (i\sigma_2\eta_{\beta}^*)
-m_{\beta\beta} (-i\sigma_2\xi_{\beta}^*)-m_{\alpha\beta} (-i\sigma_2\xi_{\alpha}^*)=0, \\
&&i\partial_0 (-i\sigma_2\xi_{\alpha}^*)-i\sigma_i\partial_i (-i\sigma_2\xi_{\alpha}^*)
-m_{\alpha\alpha}^* (i\sigma_2\eta_{\alpha}^*)-m_{\alpha\beta}^* (i\sigma_2\eta_{\beta}^*)=0, \\
&&i\partial_0 (-i\sigma_2\xi_{\beta}^*)-i\sigma_i\partial_i (-i\sigma_2\xi_{\beta}^*)
-m_{\beta\beta}^* (i\sigma_2\eta_{\beta}^*)-m_{\beta\alpha}^* (i\sigma_2\eta_{\alpha}^*)=0. \label{D.E.anti-nu4}
\end{eqnarray}
Here, we take the complex conjugate of (\ref{etaxia}) and (\ref{etaxib}), 
\begin{eqnarray}
&&\hspace{-0.5cm}\eta_{\alpha}^*(x,t)=e^{-i\vec{p}\cdot \vec{x}}\eta_{\alpha}^*(t)
=e^{-i\vec{p}\cdot \vec{x}}\left(\begin{array}{c}\nu_{\alpha L}^{*\prime} \\ \nu_{\alpha L}^*\end{array}\right), \quad 
\xi_{\alpha}^*(x,t)=e^{-i\vec{p}\cdot \vec{x}}\xi_{\alpha}^*(t)
=e^{-i\vec{p}\cdot \vec{x}}\left(\begin{array}{c}\nu_{\alpha R}^{*\prime} \\ \nu_{\alpha R}^*\end{array}\right), \\
&&\hspace{-0.5cm}\eta_{\beta}^*(x,t)=e^{-i\vec{p}\cdot \vec{x}}\eta_{\beta}^*(t)
=e^{-i\vec{p}\cdot \vec{x}}\left(\begin{array}{c}\nu_{\beta L}^{*\prime} \\ \nu_{\beta L}^*\end{array}\right), \quad
\xi_{\beta}^*(x,t)=e^{-i\vec{p}\cdot \vec{x}}\xi_{\beta}^*(t)
=e^{-i\vec{p}\cdot \vec{x}}\left(\begin{array}{c}\nu_{\beta R}^{*\prime} \\ \nu_{\beta R}^*\end{array}\right), 
\end{eqnarray}
and choose $\vec{p}=(0,0,p)$, the equations (\ref{D.E.anti-nu1})-(\ref{D.E.anti-nu4}) 
are rewritten by the matrix form, 
\begin{eqnarray}
&&\hspace{-1cm}i\partial_0 \left(\begin{array}{c}\nu_{\alpha L}^{c} \\ \nu_{\alpha L}^{c\prime}\end{array}\right)
+p\left(\begin{array}{c}\nu_{\alpha L}^{c} \\ -\nu_{\alpha L}^{c\prime}\end{array}\right)
-m_{\alpha\alpha} \left(\begin{array}{c}\nu_{\alpha R}^{c} \\ \nu_{\alpha R}^{c\prime}\end{array}\right)
-m_{\beta\alpha} \left(\begin{array}{c}\nu_{\beta R}^{c} \\ \nu_{\beta R}^{c\prime}\end{array}\right)=0, \\
&&\hspace{-1cm}i\partial_0 \left(\begin{array}{c}\nu_{\beta L}^{c} \\ \nu_{\beta L}^{c\prime}\end{array}\right)
+p\left(\begin{array}{c}\nu_{\beta L}^{c} \\ -\nu_{\beta L}^{c\prime}\end{array}\right)
-m_{\beta\beta} \left(\begin{array}{c}\nu_{\beta R}^{c} \\ \nu_{\beta R}^{c\prime}\end{array}\right)
-m_{\alpha\beta} \left(\begin{array}{c}\nu_{\alpha R}^{c} \\ \nu_{\alpha R}^{c\prime}\end{array}\right)=0, \\
&&\hspace{-1cm}i\partial_0 \left(\begin{array}{c}\nu_{\alpha R}^{c} \\ \nu_{\alpha R}^{c\prime}\end{array}\right)
-p\left(\begin{array}{c}\nu_{\alpha R}^{c} \\ -\nu_{\alpha R}^{c\prime}\end{array}\right)
-m_{\alpha\alpha}^* \left(\begin{array}{c}\nu_{\alpha L}^{c} \\ \nu_{\alpha L}^{c\prime}\end{array}\right)
-m_{\alpha\beta}^* \left(\begin{array}{c}\nu_{\beta L}^{c} \\ \nu_{\beta L}^{c\prime}\end{array}\right)=0, \\
&&\hspace{-1cm}i\partial_0 \left(\begin{array}{c}\nu_{\beta R}^{c} \\ \nu_{\beta R}^{c\prime}\end{array}\right)
-p\left(\begin{array}{c}\nu_{\beta R}^{c} \\ -\nu_{\beta R}^{c\prime}\end{array}\right)
-m_{\beta\beta}^* \left(\begin{array}{c}\nu_{\beta L}^{c} \\ \nu_{\beta L}^{c\prime}\end{array}\right)
-m_{\beta\alpha}^* \left(\begin{array}{c}\nu_{\alpha L}^{c} \\ \nu_{\alpha L}^{c\prime}\end{array}\right)=0. 
\end{eqnarray}
Equations obtained above can be combined to one matrix form,
\begin{eqnarray}
i\frac{d}{dt}\left(\begin{array}{c}
\nu_{\alpha R}^{c\prime} \\ \nu_{\beta R}^{c\prime} \\ \nu_{\alpha L}^{c\prime} \\ \nu_{\beta L}^{c\prime} \\
\nu_{\alpha R}^c \\ \nu_{\beta R}^c \\ \nu_{\alpha L}^c \\ \nu_{\beta L}^c 
\end{array}\right)
=\left(\begin{array}{cccc|cccc}
-p & 0 & m_{\alpha\alpha}^* & m_{\alpha\beta}^* & 0 & 0 & 0 & 0 \\ 
0 & -p & m_{\beta\alpha}^* & m_{\beta\beta}^* & 0 & 0 & 0 & 0 \\
m_{\alpha\alpha} & m_{\beta\alpha} & p & 0 & 0 & 0 & 0 & 0 \\ 
m_{\alpha\beta} & m_{\beta\beta} & 0 & p & 0 & 0 & 0 & 0 \\
\hline
0 & 0 & 0 & 0 & p & 0 & m_{\alpha\alpha}^* & m_{\alpha\beta}^* \\
0 & 0 & 0 & 0 & 0 & p & m_{\beta\alpha}^* & m_{\beta\beta}^* \\
0 & 0 & 0 & 0 & m_{\alpha\alpha} & m_{\beta\alpha} & -p & 0 \\
0 & 0 & 0 & 0 & m_{\alpha\beta} & m_{\beta\beta} & 0 & -p 
\end{array}\right)\left(\begin{array}{c}
\nu_{\alpha R}^{c\prime} \\ \nu_{\beta R}^{c\prime} \\ \nu_{\alpha L}^{c\prime} \\ \nu_{\beta L}^{c\prime} \\
\nu_{\alpha R}^c \\ \nu_{\beta R}^c \\ \nu_{\alpha L}^c \\ \nu_{\beta L}^c 
\end{array}\right).
\end{eqnarray}
Note that this equation is for the anti-neutrinos with negative momentum, $-p$.
Changing the sign of momentum in order to calculate the oscillation probabilities with $p$, 
the above equation is rewritten as 
\begin{eqnarray}
i\frac{d}{dt}\left(\begin{array}{c}
\nu_{\alpha R}^{c\prime} \\ \nu_{\beta R}^{c\prime} \\ \nu_{\alpha L}^{c\prime} \\ \nu_{\beta L}^{c\prime} \\
\nu_{\alpha R}^c \\ \nu_{\beta R}^c \\ \nu_{\alpha L}^c \\ \nu_{\beta L}^c 
\end{array}\right)
=\left(\begin{array}{cccc|cccc}
p & 0 & m_{\alpha\alpha}^* & m_{\alpha\beta}^* & 0 & 0 & 0 & 0 \\ 
0 & p & m_{\beta\alpha}^* & m_{\beta\beta}^* & 0 & 0 & 0 & 0 \\
m_{\alpha\alpha} & m_{\beta\alpha} & -p & 0 & 0 & 0 & 0 & 0 \\ 
m_{\alpha\beta} & m_{\beta\beta} & 0 & -p & 0 & 0 & 0 & 0 \\
\hline
0 & 0 & 0 & 0 & -p & 0 & m_{\alpha\alpha}^* & m_{\alpha\beta}^* \\
0 & 0 & 0 & 0 & 0 & -p & m_{\beta\alpha}^* & m_{\beta\beta}^* \\
0 & 0 & 0 & 0 & m_{\alpha\alpha} & m_{\beta\alpha} & p & 0 \\
0 & 0 & 0 & 0 & m_{\alpha\beta} & m_{\beta\beta} & 0 & p 
\end{array}\right)\left(\begin{array}{c}
\nu_{\alpha R}^{c\prime} \\ \nu_{\beta R}^{c\prime} \\ \nu_{\alpha L}^{c\prime} \\ \nu_{\beta L}^{c\prime} \\
\nu_{\alpha R}^c \\ \nu_{\beta R}^c \\ \nu_{\alpha L}^c \\ \nu_{\beta L}^c 
\end{array}\right).
\end{eqnarray}
Comparing this to (\ref{8-8-nu-matrix}), one can see that the matrix for anti-neutrinos 
can be obtained by the replacement as follows; 
\begin{eqnarray}
\left(\begin{array}{c}\nu_{\alpha R} \\ \nu_{\beta R} \\ \nu_{\alpha L} \\ \nu_{\beta L} \end{array}\right)
\to 
\left(\begin{array}{c}\nu_{\alpha R}^{c} \\ \nu_{\beta R}^{c} \\ \nu_{\alpha L}^{c} \\ \nu_{\beta L}^{c}
 \end{array}\right), 
\qquad 
\left(\begin{array}{c}\nu_{\alpha R}^{\prime} \\ \nu_{\beta R}^{\prime} \\ \nu_{\alpha L}^{\prime} \\ \nu_{\beta L}^{\prime} \end{array}\right)
\to 
\left(\begin{array}{c} \nu_{\alpha R}^{c\prime} \\ \nu_{\beta R}^{c\prime} \\ 
\nu_{\alpha L}^{c\prime} \\ \nu_{\beta L}^{c\prime}\end{array}\right), 
\qquad m \to m^*.
\end{eqnarray}
As seen from the above replacement, the CP conjugate state of $\nu_L$ is considered to be $\nu_L^{c}$.
In the anti-neutrino case, the unitary matrices $U$ and $V$ diagonalizing the mass part are changed 
into $U^*$ and $V^*$ according to the complexation of the mass terms.
Therefore, the sign of the phase $\phi$ is reversed.
So, the oscillation probabilities for anti-neutrinos are given by 
\begin{eqnarray}
&&\hspace{-1cm}P(\nu_{\alpha L}^{c}\to\nu_{\alpha L}^{c})=P(\nu_{\alpha L}\to\nu_{\alpha L})\nonumber \\
&&=1-4s_L^2c_L^2\sin^2 \frac{(E_2-E_1)t}{2} \nonumber \\
&&\hspace{-1cm}-\left[c_L^4\cdot \frac{m_1^2}{E_1^2}\sin^2 (E_1t)
+s_L^4\cdot \frac{m_2^2}{E_2^2}\sin^2 (E_2t)+2s_L^2c_L^2\left(1-\frac{p^2}{E_1E_2}\right)
\sin (E_1t)\sin (E_2t)\right], \\
&&\hspace{-1cm}P(\nu_{\alpha L}^{c}\to\nu_{\beta L}^{c})=P(\nu_{\alpha L}\to\nu_{\beta L})\nonumber \\
&&=4s_L^2c_L^2\sin^2 \frac{(E_2-E_1)t}{2} \nonumber \\
&&-s_L^2c_L^2\left[\frac{m_1^2}{E_1^2}\sin^2 (E_1t)+\frac{m_2^2}{E_2^2}\sin^2 (E_2t)
-2\left(1-\frac{p^2}{E_1E_2}\right)\sin (E_1t)\sin (E_2t)\right], \\
&&\hspace{-1cm}P(\nu_{\alpha L}^{c}\to\nu_{\alpha R}^{c})=P(\nu_{\alpha L}\to\nu_{\alpha R})\nonumber \\
&&=c_L^2c_R^2\frac{m_1^2}{E_1^2}\sin^2 (E_1t)+s_L^2s_R^2\frac{m_2^2}{E_2^2}\sin^2 (E_2t)\nonumber \\
&&+2s_Ls_Rc_Lc_R\cos (\phi_R-\phi_L)\frac{m_1m_2}{E_1E_2}\sin (E_1t)\sin (E_2t), \\
&&\hspace{-1cm}P(\nu_{\alpha L}^{c}\to\nu_{\beta R}^{c})=P(\nu_{\alpha L}\to\nu_{\beta R})\nonumber \\
&&=c_L^2s_R^2\frac{m_1^2}{E_1^2}\sin^2 (E_1t)
+s_L^2c_R^2\frac{m_2^2}{E_2^2}\sin^2 (E_2t)\nonumber \\
&&-2s_Ls_Rc_Lc_R\cos (\phi_R-\phi_L)\frac{m_1m_2}{E_1E_2}\sin (E_1t)\sin (E_2t). 
\end{eqnarray}
One can see that the probabilities of anti-neutrinos are the same as those for neutrinos in two generation scheme.
The point is that $U$ and $V$ become complex conjugates in the case of anti-neutrinos.
This is the origin of the difference between neutrinos and anti-neutrinos 
in more than three generations.

\section{Summary} 

To summarize this paper, we have rigorously formalized the theory of neutrino oscillations using the Dirac equation. We focused on the case of two-generation Dirac neutrinos in vacuum. Our approach provides a unified understanding of neutrino oscillations with and without chirality-flip, which were previously discussed separately in the literature. We have also demonstrated that unitarity holds only after considering both types of oscillations.

The relativistic method developed here can be extended to oscillations involving 
$n$ generations, as well as to scenarios incorporating matter effects and magnetic fields. 
We discovered the emergence of two new features in the oscillation probabilities: one is that the oscillation probabilities depend on the absolute mass of the neutrinos, while the other is the appearance of a new CP phase. This discovery holds even when considering only two generations.
This phase is comprehensive, encompassing the Majorana phase 
\cite{Bernabeu1983,Majorana-phase,Majorana-phase2,Majorana-phase3}.

The first new feature suggests that neutrino oscillation experiments could, in principle, measure not only mass squared differences but also the absolute mass. The second feature is that the effect of the new CP phase may be detectable if there are interactions capable of distinguishing the flavor of right-handed neutrinos in physics beyond the Standard Model. These new features could be comparatively significant in neutrino oscillations at atomic scales, such as in $0\nu\beta\beta$ decay.

Conversely, in previous experiments, the contributions from these new effects were negligible, and no contradictions were found. In summary, we have successfully extended the theory of neutrino oscillations in a natural and relativistic manner.

\section*{Note added in proof} 

We learned that the literature \cite{Li2024} was published in 2024 after we submitted our paper to arXiv in 2021. The authors of this paper derived the oscillation probabilities for Majorana neutrinos using the same method as ours. The vacuum oscillation probabilities, Eq.(19), they derived are the same as the results Eqs.(84)-(87) in our paper \cite{KT2}, both in the presence and absence of chirality-flips. Furthermore, they extended this method to cases with matter effects, pointing out that new resonance effects emerge related to helicity states, which are distinct from the usual MSW effect. They also noted that in cases where non-relativistic effects become significant, such as with cosmic background neutrinos, this matter effect may provide a possibility for measuring the Majorana CP phase.


\begin{thebibliography}{99}

\bibitem{Pontecorvo} B. Pontecorvo, 
\emph{Mesonium and anti-mesonium},
 \emph{Sov. Phys. JETP} {\bf 6}, (1957) 429.

\bibitem{MNS} Z. Maki, M. Nakagawa, and S. Sakata,
\emph{Remarks on the unified model of elementary particles},
 \emph{Prog. Theor. Phys.} {\bf 28}, (1962) 870.

\bibitem{3-gene}
V. D. Barger, K. Whisnant, R.J.N. Phillips, 
 \emph{CP Violation in Three Neutrino Oscillations}, 
 \emph{Phys.Rev.Lett.} {\bf 45} (1980) 2084.

\bibitem{MSW1}
L. Wolfenstein,  \emph{Neutrino oscillations in matter}, \emph{Phys. Rev.} {\bf D17} (1978) 2369.

\bibitem{MSW2}
S. P. Mikheev and A. Yu. Smirnov,
\emph{Resonant amplification of neutrino oscillations in matter and solar neutrino
spectroscopy}. \emph{Nuovo Cim. C}, {\bf 9}, (1986) 17.

\bibitem{mag}
L.B. Okun, M.B. Voloshin, M.I. Vysotsky, 
 \emph{Electromagnetic Properties of Neutrino and Possible Semiannual Variation Cycle of the Solar Neutrino Flux}, 
 \emph{Sov.J.Nucl.Phys.} {\bf 44} (1986) 440,  \emph{Yad.Fiz.} {\bf 44} (1986) 677.


\bibitem{grossman}
Y. Grossman,
\emph{Nonstandard neutrino interactions and neutrino oscillation experiments},
\emph{Phys.Lett.B} {\bf 359} (1995) 141
[hep-ph/9507344].

\bibitem{NSI1}
T. Ohlsson. 
\emph{Status of non-standard neutrino interactions}. 
\emph{Rept. Prog. Phys.}, {\bf 76} (2013) 044201,
[arXiv:1209.2710 [hep-ph]].


\bibitem{NSI2}
O. G. Miranda and H. Nunokawa. 
\emph{Non standard neutrino interactions: current status and future prospects}. 
\emph{New J. Phys.}, {\bf 17} (2015) (9) 095002,
[arXiv:1505.06254 [hep-ph]].

\bibitem{NSI3}
Y. Farzan and M. Tortola. 
\emph{Neutrino oscillations and Non-Standard Interactions}. 
\emph{Front.in Phys.}, {\bf 6} (2018) 10,
[arXiv:1710.09360 [hep-ph]].

\bibitem{NSI4}
P. S. Bhupal Dev et al. 
\emph{Neutrino Non-Standard Interactions}: 
A Status Report. In NTN Workshop on Neutrino
Non-Standard Interactions St Louis, MO, USA, May 29-
31, 2019, 2019, 
[arXiv:1907.00991 [hep-ph]].

\bibitem{SKatmos}
 Y.~Fukuda {\it et al.} [Super-Kamiokande Collaboration],
\emph{Evidence for oscillation of atmospheric neutrinos}.
\emph{Phys.Rev.Lett.} {\bf 81} (1998) 1562
[hep-ex/9807003].


\bibitem{SK}
S.~Moriyama,  \emph{New atmospheric and solar results from Super-Kamiokande}, Talk at XXVII International Conference on Neutrino Physics and Astrophysics, London, 4-9 July, 2016. 

\bibitem{SNO}
  B.~Aharmim {\it et al.} [SNO Collaboration],
 \emph{Combined Analysis of all Three Phases of Solar Neutrino Data from the Sudbury Neutrino Observatory},
   \emph{Phys. Rev. C} {\bf 88} (2013) 025501
  [arXiv:1109.0763 [nucl-ex]].

\bibitem{SK2}
  K.~Abe {\it et al.} [Super-Kamiokande Collaboration],
  \emph{Solar Neutrino Measurements in Super-Kamiokande-IV},
   \emph{Phys. Rev. D} {\bf 94} (2016) no.5,  052010
  [arXiv:1606.07538 [hep-ex]].

\bibitem{T2K}
  K.~Abe {\it et al.} [T2K Collaboration],
 \emph{Indication of Electron Neutrino Appearance from an Accelerator-produced Off-axis Muon Neutrino Beam},
   \emph{Phys. Rev. Lett.}  {\bf 107} (2011) 041801
  [arXiv:1106.2822 [hep-ex]].


\bibitem{MINOS}
  P.~Adamson {\it et al.} [MINOS Collaboration],
 \emph{Combined analysis of $\nu_{\mu}$ disappearance and $\nu_{\mu} \rightarrow \nu_{e}$ appearance in MINOS using accelerator and atmospheric neutrinos},
   \emph{Phys. Rev. Lett.}  {\bf 112} (2014) 191801.
  [arXiv:1403.0867 [hep-ex]].

\bibitem{NOvA}
M. A. Acero et al.,  [NOvA Collaboration], 
 \emph{First Measurement of Neutrino Oscillation Parameters using Neutrinos and Antineutrinos by NOvA}, 
 \emph{Phys.Rev.Lett.} {\bf 123} (2019) no.15, 151803, 
[arXiv:1906.04907 [hep-ex]].

\bibitem{KamLAND}
  A.~Gando {\it et al.} [KamLAND Collaboration],
 \emph{Reactor On-Off Antineutrino Measurement with KamLAND},
   \emph{Phys. Rev. D} {\bf 88} (2013) 3,  033001
   [arXiv:1303.4667 [hep-ex]].

\bibitem{DayaBay}
  F.~P.~An {\it et al.} [Daya Bay Collaboration],
 \emph{Measurement of electron antineutrino oscillation based on 1230 days of operation of the Daya Bay experiment},
  arXiv:1610.04802 [hep-ex].

\bibitem{RENO}
  J.~H.~Choi {\it et al.} [RENO Collaboration],
  \emph{Observation of Energy and Baseline Dependent Reactor Antineutrino Disappearance in the RENO Experiment},
   \emph{Phys. Rev. Lett.}  {\bf 116} (2016) no.21,  211801
  [arXiv:1511.05849 [hep-ex]].

\bibitem{DoubleChooz}
  Y.~Abe {\it et al.} [Double Chooz Collaboration],
 \emph{Improved measurements of the neutrino mixing angle $\theta_{13}$ with the Double Chooz detector},
   \emph{JHEP} {\bf 1410} (2014) 086
   Erratum: [ \emph{JHEP} {\bf 1502} (2015) 074]
  [arXiv:1406.7763 [hep-ex]].

\bibitem{Dirac CP}
  K.~Abe {\it et al.} [T2K Collaboration],
 \emph{Constraint on the Matter-Antimatter Symmetry-Violating Phase in Neutrino Oscillations},
   \emph{Nature} {\bf 580} (2020) no.7803, 339
  [arXiv:1910.03887 [hep-ex]].

\bibitem{HK}
  K.~Abe {\it et al.} [Hyper-Kamiokande Proto-Collaboration],
  \emph{Physics potential of a long-baseline neutrino oscillation experiment using a J-PARC neutrino beam and Hyper-Kamiokande},
   \emph{PTEP} {\bf 2015} (2015) 053C02
    [arXiv:1502.05199 [hep-ex]].

\bibitem{DUNE}
  R.~Acciarri {\it et al.} [DUNE Collaboration],
  \emph{Long-Baseline Neutrino Facility (LBNF) and Deep Underground Neutrino Experiment (DUNE) : Volume 2: The Physics Program for DUNE at LBNF},
  arXiv:1512.06148 [physics.ins-det].

\bibitem{Zaglauer} H. W. Zaglauer and K. H. Schwarzer,
\emph{The Mixing Angles in Matter for Three Generations of Neutrinos and the Msw Mechanism},
 \emph{Z. Phys. C} {\bf 40} (1988) 273. 

\bibitem{Ohlsson} T. Ohlsson and H. Snellman, 
\emph{Three flavor neutrino oscillations in matter},
 \emph{J. Math. Phys.} {\bf 41} (2000) 2768;  \emph{Phys. Lett. B} {\bf 474} (2000) 153.


\bibitem{KTY1}
K. Kimura, A. Takamura and H. Yokomakura,
\emph{Exact formula of probability and CP violation for neutrino oscillations in matter},
\emph{Phys. Lett. B} {\bf 537} (2002) 86.


\bibitem{KTY2}
K. Kimura, A. Takamura and H. Yokomakura,
\emph{Exact formulas and simple CP dependence of neutrino oscillation probabilities in matter with constant density}
 \emph{Phys. Rev. D} {\bf 66} (2002) 073005.

\bibitem{Yokomakura0207} 
 H. Yokomakura, K. Kimura and A. Takamura, 
\emph{Overall feature of CP dependence for neutrino oscillation probability in arbitrary matter profile},
 \emph{Phys. Lett. B} {\bf 544} (2002) 286.

\bibitem{Yasuda} 
 O. Yasuda, 
\emph{Nonadiabatic contributions to the neutrino oscillation probability and the formalism by Kimura, Takamura, and Yokomakura},
  \emph{Phys. Rev. D} {\bf 89} (2014) 093023.



\bibitem{Fukugita-note}
M. Fukugita and T. Yanagida,  \emph{Physics of neutrinos}, (1993) p101.

\bibitem{Majorana} 
E. Majorana,  
\emph{Teoria simmetrica dell'elettrone e del positrone},
\emph{Nuovo Cimento} {\bf 14} (1937) 171.

\bibitem{Bahcall1978}
J. Bahcall and H. Primakoff, 
\emph{Neutrino-antineutrinos Oscillations},
 \emph{Phys. Rev. D} {\bf 18} (1978) 3463.

\bibitem{Valle1981} 
J. Schechter and J.W.F. Valle, 
\emph{Neutrino Oscillation Thought Experiment},
 \emph{Phys. Rev. D} {\bf 23} (1981) 1666. 

\bibitem{Li1982} 
L.F. Li and F. Wilczek, 
\emph{PHYSICAL PROCESSES INVOLVING MAJORANA NEUTRINOS},
 \emph{Phys. Rev. D} {\bf 25} (1982) 143.

\bibitem{Bernabeu1983} 
J. Bernabeu and P. Pascual, 
\emph{CP Properties of the Leptonic Sector for Majorana Neutrinos},
\emph{Nucl. Phys. B} {\bf 228} (1983) 21.

\bibitem{Gouvea2003}
A. de Gouvea, B. Kayser, and R.N. Mohapatra,
\emph{Manifest CP Violation from Majorana Phases},
 \emph{Phys. Rev. D} {\bf 67} (2003) 053004.

\bibitem{Xing2013} 
Z.Z. Xing, 
\emph{Properties of CP Violation in Neutrino-Antineutrino Oscillations},
 \emph{Phys. Rev. D} {\bf 87} (2013) 053019.

\bibitem{BBB2020}
V. A. S. V. Bittencourt, A. E. Bernardini and M. Blasone, 
\emph{Chiral oscillations in the non-relativistic regime},
arXiv:2009.00084 [hep-ph].

\bibitem{Ge2020}
S. F. Ge and P. Pasquini, 
\emph{Parity Violation and Chiral Oscillation of Cosmological Relic Neutrinos},
\emph{Phys. Lett. B} {\bf 811} (2020) 135961, arXiv:2009.01684 [hep-ph].



\bibitem{Borah2017}
D. Borah and A. Dasgupta, 
\emph{Naturally Light Dirac Neutrino in Left-Right Symmetric Model},
\emph{JCAP} {\bf 06} (2017) 003, arXiv:1702.02877 [hep-ph].

\bibitem{Dvornikov}
M. Dvornikov, 
 \emph{Field theory description of neutrino oscillations}, 
Neutrinos: Properties, Sources and Detection, ed. by J.P.Greene. (Nova Science Publishers, New York, 2011, p. 23-90,
arXiv:1011.4300. 

\bibitem{Bernardini2004}
A. E. Bernardini and S. D. Leo,
\emph{Dirac Spinors and Flavor Oscillations}
\emph{Eur.Phys.J.C} {\bf 37} (2004) 471,
hep-ph/0411153.

\bibitem{Bernardini2005}
A. E. Bernardini and S. D. Leo,
\emph{Flavor and chiral oscillations with Dirac wave packets}
\emph{Phys.Rev. D} {\bf 71} (2005) 076008,
hep-ph/0504239.

\bibitem{Nishi}
A. E. Bernardini, M. M. Guzzo and C. C. Nishi, 
\emph{Quantum flavor oscillations extended to the Dirac theory}, 
\emph{Fortsch.Phys.} {\bf 59} (2011) 372. 
arXiv:1004.0734. 


\bibitem{Majorana-phase} 
S.M. Bilenky, J. Hosek, and S.T. Petcov,
\emph{On Oscillations of Neutrinos with Dirac and Majorana Masses},
 \emph{Phys. Lett. B} {\bf 94} (1980) 495.
 
\bibitem{Majorana-phase2} 
J. Schechter and J.W.F. Valle,
\emph{Neutrino Masses in SU(2) x U(1) Theories},
 \emph{Phys. Rev. D} {\bf 22} (1980) 2227.
 
\bibitem{Majorana-phase3} 
M. Doi, T. Kotani, H. Nishiura, K. Okuda, and E. Takasugi,  
 \emph{CP Violation in Majorana Neutrinos},
 \emph{Phys. Lett. B} {\bf 102} (1981) 323.

\bibitem{Li2024}
Ming-Wei Li, Zhong-Lv Huang and Xiao-Gang He, 
\emph{Parity Violation and Chiral Oscillation of Cosmological Relic Neutrinos},
\emph{Phys. Lett. B} {\bf 855} (2024) 138778, arXiv:2307.12561 [hep-ph].


\bibitem{KT2}
K. Kimura and A. Takamura,
\emph{Unification of Neutrino-Neutrino and Neutrino-Antineutrino Oscillations},
arXiv:2101.04509.




\end{thebibliography}
\end{document}